\documentclass[a4paper,twocolumn,11pt]{quantumarticle}
\pdfoutput=1
\usepackage[utf8]{inputenc}
\usepackage[english]{babel}
\usepackage[T1]{fontenc}
\usepackage{amsmath}
\usepackage{hyperref}
\usepackage{url}

\usepackage[ruled,resetcount]{algorithm2e}

\newcommand{\kro}{r}
\newcommand{\kroF}{R}
\newcommand{\kroM}{r_{\text{max}}}

\newcommand{\aveErr}{\xi}
\usepackage{tikz}
\usepackage{lipsum}
\usepackage{braket}
\usepackage[normalem]{ulem}

\SetKw{KwParams}{Parameters}
\SetKw{KwSub}{Subroutines}
\SetKw{KwInit}{Initialisation}
\SetKw{KwBegin}{Begin}
\SetKwProg{Fn}{}{}{end}

\SetCommentSty{mycommfont}

\begin{document}

\title{Partitioned Quantum Subspace Expansion}

\author{Tom O'Leary}
\affiliation{Clarendon Laboratory, University of Oxford, Parks Road, Oxford OX1
3PU, United Kingdom}
\affiliation{Theoretical Division, Los Alamos National Laboratory, Los Alamos, NM, USA}
\email{thomas.oleary@lmh.ox.ac.uk}
\orcid{0009-0003-2065-1695}
\author{Lewis W. Anderson}
\orcid{0000-0003-0269-3237}
\affiliation{Clarendon Laboratory, University of Oxford, Parks Road, Oxford OX1
3PU, United Kingdom}
\author{Dieter Jaksch}
\affiliation{Clarendon Laboratory, University of Oxford, Parks Road, Oxford OX1
3PU, United Kingdom}
\affiliation{Institut für Quantenphysik, Universit\"{a}t Hamburg, Hamburg, Germany}
\orcid{0000-0002-9704-3941}
\author{Martin Kiffner}
\affiliation{Clarendon Laboratory, University of Oxford, Parks Road, Oxford OX1
3PU, United Kingdom}
\affiliation{PlanQC GmbH, Lichtenbergstr. 8, 85748 Garching, Germany}
\orcid{0000-0002-8321-6768}
\maketitle

\begin{abstract}
We present an iterative generalisation of the quantum subspace expansion algorithm used with a Krylov basis. The iterative construction connects a sequence of subspaces via their lowest energy states. Diagonalising a Hamiltonian in a given Krylov subspace requires the same quantum resources in both the single step and sequential cases. We propose a variance-based criterion for determining a good iterative sequence and provide numerical evidence that these good sequences display improved numerical stability over a single step in the presence of finite sampling noise. Implementing the generalisation requires additional classical processing with a polynomial overhead in the subspace dimension. By exchanging quantum circuit depth for additional measurements the quantum subspace expansion algorithm appears to be an approach suited to near term or early error-corrected quantum hardware. Our work suggests that the numerical instability limiting the accuracy of this approach can be substantially alleviated in a parameter-free way.
\end{abstract}

\section{Introduction}
Finding accurate ground state energies of complex quantum systems
is of paramount importance for fundamental research and applied chemistry applications.
However, the exponential scaling of the Hilbert space dimension with system size means that
many problems of interest cannot be solved by classical computers. On the other hand,
quantum computers can  in principle address this problem by storing and processing
complex quantum states with $2^n$ degrees of freedom  using $n$ qubits.

\begin{figure}
    \centering
    \includegraphics[trim=4cm 2cm 4cm 4cm, width=0.5\textwidth]{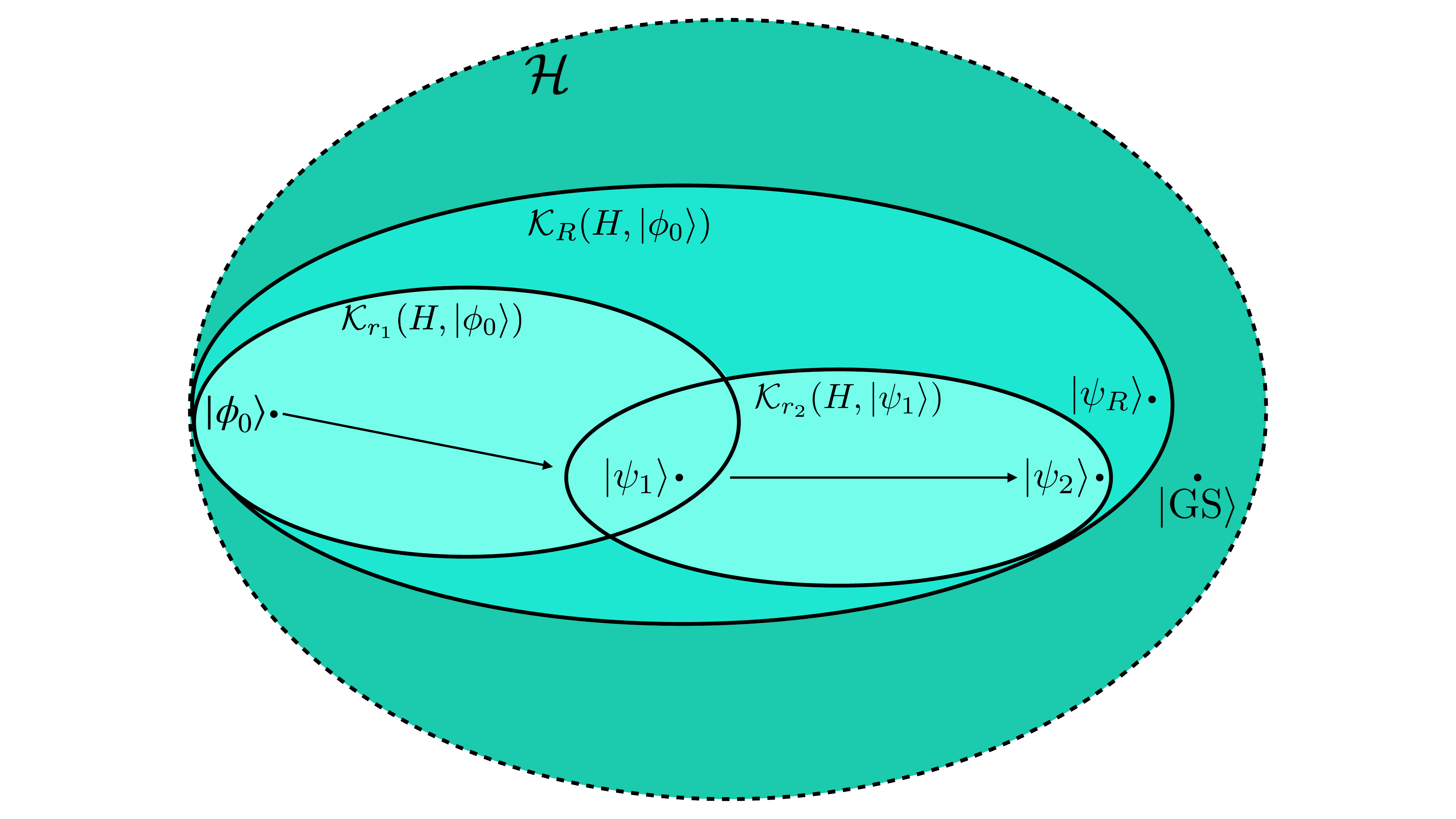}
    \caption{
    Illustration of a Partitioned Quantum Subspace Expansion (PQSE)
    for $P=2$ partitions. $\ket{\text{GS}}$ denotes the ground state of 
    a Hamiltonian $H$. When diagonalising $H$ in the order $R$ 
    Krylov subspace $\mathcal{K}_R(H,\ket{\phi_0})$, numerical instabilities can prohibit access to the ground state estimate $\ket{\psi_R}$. PQSE can avoid this by carrying out two sequential 
    steps: First we diagonalise $H$ in $\mathcal{K}_{r_1}(H,\ket{\phi_0})$ with $r_1<R$. 
    We then use the lowest energy state $\ket{\psi_1}$ of $H$ in 
    $\mathcal{K}_{r_1}(H,\ket{\phi_0})$ to construct a second subspace $\mathcal{K}_{r_2}(H,\ket{\psi_1})$, where  $r_1 + r_2 - 1= R$, with lowest energy state $\ket{\psi_2}$. For certain values of $\{r_1, r_2\}$, the ground state estimate $\ket{\psi_2}$ avoids the numerical instabilities which prohibit access to $\ket{\psi_R}$.}
    \label{fig:pqse_schem}
\end{figure}

In the era of noisy quantum computing without error correction, hybrid quantum-classical algorithms
have been proposed for a range of applications, for example: efficiently finding the ground state energy of non-linear mean-field models \cite{lubaschVariationalQuantumAlgorithms2020, kyriienkoSolvingNonlinearDifferential2021}; dynamical mean field theory algorithms for strongly correlated systems \cite{kreulaNonlinearQuantumclassicalScheme2016, kreulaFewqubitQuantumclassicalSimulation2016}; and optimization problems from adjacent fields \cite{andersonCoarseGrainedIntermolecular2022, jaderbergRecompilationenhancedSimulationElectron2022, jakschVariationalQuantumAlgorithms2023, mottaSubspaceMethodsElectronic2024, huangQuantumAdvantageLearning2022, cerezoVariationalQuantumFidelity2020,
cerezoVariationalQuantumAlgorithms2021}. Two general approaches underlying several of these algorithms are
the Variational Quantum Eigensolver (VQE)~\cite{peruzzoVariationalEigenvalueSolver2014} and
Quantum Subspace Expansion (QSE)~\cite{stairMultireferenceQuantumKrylov2020,
mccleanHybridQuantumClassicalHierarchy2017,
mccleanDecodingQuantumErrors2020}  methods. 
QSE is a quantum-classical hybrid algorithm like VQE, but the classical 
optimisation problem is mapped to a generalised eigenvalue problem, which has various 
practical advantages compared with the highly non-linear optimisation required in VQE. In 
addition, it has been shown~\cite{yoshiokaGeneralizedQuantumSubspace2022a, yangDualGSEResourceefficientGeneralized2024, getelinaQuantumSubspaceExpansion2024}  that QSE can be 
combined with powerful error mitigation techniques~\cite{hugginsVirtualDistillationQuantum2021, koczorExponentialErrorSuppression2021, temmeErrorMitigationShortdepth2017}, which makes this 
approach very attractive for 
noisy quantum devices.

The crucial task in QSE consists of generating a subspace with a lowest energy state that has properties which are sufficiently close to those of the true ground state. Most schemes for 
generating this subspace aim to create a Krylov subspace due to the excellent performance of this approach in classical algorithms 
\cite{saadRatesConvergenceLanczos1980}.
A Krylov subspace is defined by a set of non-orthogonal basis states constructed through application of powers of the Hamiltonian to a reference state.
A Krylov subspace can also be generated by performing real~\cite{stairMultireferenceQuantumKrylov2020, parrishQuantumFilterDiagonalization2019, 
klymkoRealTimeEvolutionUltracompact2022} and imaginary~\cite{mottaDeterminingEigenstatesThermal2020} time evolution, by forming a superposition of 
real time evolutions~\cite{sekiQuantumPowerMethod2021} (which can remove the need for the Hadamard test~\cite{cortesQuantumKrylovSubspace2022}), by decomposing Hamiltonian powers into products of unitary operators to form a fine grained Krylov subspace~\cite{bhartiIterativeQuantumassistedEigensolver2021}, or by generating Gaussian powers of the Hamiltonian~\cite{leeSamplingErrorAnalysis2024}.
Recently, it has been shown~\cite{kirbyExactEfficientLanczos2023} that QSE with an exact Krylov basis can be realized via qubitization~\cite{low2019hamiltonian} and applied to problems from condensed matter physics, quantum chemistry and quantum field theory \cite{kirbyExactEfficientLanczos2023, 
andersonSolvingLatticeGauge2025}. 

An advantageous feature of QSE with a Krylov basis is that the accuracy of the ground state energy increases exponentially with the number of basis 
states~\cite{saadRatesConvergenceLanczos1980,paige1971computation}. However, in practice the basis size cannot be increased arbitrarily because  higher-order 
Krylov states become almost linearly dependent due to properties of the power iteration. Consequently, the overlap and Hamiltonian matrices become 
ill-conditioned such that noise in the matrix elements measured on quantum hardware leads to numerical instability. 
It has been recently shown that  
this problem can be mitigated by projecting the generalised eigenvalue problem onto a subspace with dominant singular 
values~\cite{epperlyTheoryQuantumSubspace2022}. This thresholded QSE (TQSE) approach results in numerical stability up to a critical basis length, 
but increasing the accuracy by extending the basis size remains a desirable yet elusive capability. 
Furthermore, the effectiveness of TQSE is reliant on finding a suitable threshold parameter. This parameter will depend on properties of the target problem and noise. Although good parameters are known to exist \cite{epperlyTheoryQuantumSubspace2022, kirbyExactEfficientLanczos2023, kirbyAnalysisQuantumKrylov2024}, finding these can be challenging in practice.

Here we show that the accuracy of QSE with statistical noise can be made comparable to that achieved by TQSE in a parameter-free way.
We achieve this by introducing a Partitioned Quantum Subspace Expansion (PQSE), where one large QSE instance is partitioned into several smaller QSE problems which exhibit better conditioning. The introduced PQSE scheme is iterative, with the output of one small QSE instance serving as input to the next small QSE problem. 
The scheme is illustrated in Fig.~(\ref{fig:pqse_schem}) for the simple case of a subspace being partitioned into two smaller subspaces [a technical introduction to the figure is provided in Sec.~(\ref{PQSE})].
PQSE is intended for operation in the Krylov basis regime where numerical instability becomes an obstacle to the effectiveness of TQSE. Overall PQSE requires the same quantum resources but more classical calculations than QSE or TQSE with a fixed threshold parameter. The classical cost is determined by the choice of partitioning. By leveraging additional quantum resources, we provide a criterion for choosing an effective partitioning based on the energy variance.
We analyse two canonical examples, a spin ring with random potential and a molecular Hamiltonian. We find that the size of the Krylov basis can be substantially increased compared to standard QSE, and thus PQSE achieves substantially higher accuracy, comparable to that achieved by TQSE.

This paper is organised as follows. In Sec.~\ref{background} we introduce how a Krylov subspace is used with QSE and how thresholding can be used to recover a level of numerical stability. In Sec.~\ref{PQSE} we describe the iterative subspace construction that characterises PQSE. This is followed in Sec.~\ref{sec:num} by numerical results. In Sec.~\ref{sec:disc} we conclude with a discussion of the practical application of PQSE and further points of research, as well as a comparison with an existing iterative QSE algorithm.

\section{Methods} \label{sec:meth}
We begin with a brief review of standard QSE and its thresholded extension TQSE in Sec.~\ref{background}. The partitioned approach PQSE is introduced in Sec.~\ref{PQSE}.
\subsection{Background \label{background}}
QSE involves diagonalising an $n$ qubit Hamiltonian $H$ in a $d$ dimensional subspace spanned by typically non-orthogonal basis states. Writing
the basis states as $\{ \ket{\phi_i} \}$ for $i\in\{0,\ldots d-1\}$  we can define a trial state
\begin{align}
\ket{\psi} & = \sum\limits_{i=0}^{d-1} c_i \ket{\phi_i}\,.
\label{expansion}
\end{align}
The coefficients which define the subspace ground state can be found by minimising the Ritz coefficient
 \begin{align}
  E=\frac{\braket{\psi \lvert H \rvert \psi}}{\braket{\psi\vert\psi}}.
 \end{align}
 The optimal coefficients are then solutions to the generalised eigenvalue problem
\begin{align} \label{eq:gev}
    \mathbf{H} \mathbf{c} = E \mathbf{S} \mathbf{c}\,,
\end{align}
with matrix elements
\begin{align}
    H_{ij} & = \braket{\phi_i \lvert H \rvert \phi_j}\,, \\
S_{ij} & = \braket{\phi_i \lvert \phi_j}.
\end{align}
QSE will output a set of optimal coefficients $\mathbf{c}~=~[c_0, c_1,\dots c_{d-1}]^T$ which characterise the lowest energy combination of basis
states with respect to $H$.
The solution to Eq.~(\ref{eq:gev}) is a solution to the full eigenvalue problem of the Hamiltonian if and only if the variance of the
Hamiltonian in this state vanishes, i.e. $\text{var}(H,\ket{\psi})=0$ with

\begin{equation} \label{eq:var}
\text{var}(H,\ket{\psi}) =\frac{1}{\braket{\psi\vert\psi}}\left( \bra{\psi}H^2\ket{\psi}-\bra{\psi}H \ket{\psi}^2\right).
\end{equation}
This criterion determines a ground state solution if such a state can be expressed by the subspace basis states. 

A special choice of subspace in Eq.~(\ref{expansion}) is given by a 
Krylov subspace. A Krylov subspace of order $\kroF$, generated by an operator $A$ and starting from a reference state $\ket{\phi_0}$ is defined as

\begin{align} \label{eq:krylov_basis}
\mathcal{K}_{\kroF}(A,\ket{\phi_0}) =\text{span} \{ \ket{\phi_0}, A \ket{\phi_0} ,\ldots, A^{\kroF-1} \ket{\phi_0}\}.
\end{align}
The QSE matrices are then
\begin{subequations}
\label{eq:qse_mat_gen}
\begin{align} 
    H_{ij} & = \braket{\phi_0 \lvert (A^{i})^{\dagger} H A^{j} \rvert \phi_0}\,, \\
S_{ij} & = \braket{\phi_0 \lvert (A^{i})^{\dagger}A^{j} \rvert \phi_0}\,, \\
i,j &\in \{0,1,\ldots,\kroF-1\}.
\end{align}
\end{subequations}
Taking $A=H$ the matrices are
\begin{subequations}
\label{eq:qse_mat}
\begin{align} 
    H_{ij} & = \braket{\phi_0 \lvert H^{i + j + 1} \rvert \phi_0}\,, \\
S_{ij} & = \braket{\phi_0 \lvert H^{i + j} \rvert \phi_0}\,, \\
i,j &\in \{0,1,\ldots,\kroF-1\}.
\end{align}
\end{subequations}
With a Hamiltonian power basis these QSE matrices have a Hankel structure. This structure allows both QSE matrices to be constructed by calculating the first row and final column of
$\mathbf{H}$.

TQSE involves projecting the QSE matrices onto the eigenspace of $\mathbf{S}$ with singular values greater than a threshold $\tau$~\cite{epperlyTheoryQuantumSubspace2022}.  Concretely, if $V$ is a matrix with the eigenvectors of $\mathbf{S}$ as columns and $D$ is a diagonal matrix formed by the eigenvalues  of $\mathbf{S}$ then we use $I = \{ i : D_{ii} > \tau\}$ to redefine $V = V (:, I)$. $V$ projects onto the subspace containing eigenvalues above $\tau$. We solve $V^{\dagger} \mathbf{H} V \mathbf{c} = E V^{\dagger} \mathbf{S} V \mathbf{c}$ to find the lowest energy of the thresholded subspace. 

\subsection{Partitioned Quantum Subspace Expansion \label{PQSE}}
We illustrate the idea of our algorithm by considering  $P$ sequential QSE problems. 
We will focus on a Hamiltonian power basis i.e. $A = H$, and show how the method can be straightforwardly generalized later in the text.
A guiding schematic for $P=2$ is shown in Fig.~\ref{fig:pqse_schem}. 
Starting from an initial state $\ket{\phi_0}$, we build the subspace 
\begin{align}
 \mathcal{K}_{\kro_1}(H,\ket{\phi_0})\,,
\end{align}
 and denote its ground state by $\ket{\psi_1}$. Next we form the subspace
\begin{align}
 \mathcal{K}_{\kro_2}(H,\ket{\psi_1})\,,
\end{align}
and denote its corresponding ground state by $\ket{\psi_2}$.
The largest power of $H$ in $\ket{\psi_2}$ will be $\kro_1 + \kro_2 - 2$. 
More generally, we consider the sequence 
\begin{align} \label{eq:pqse_seq}
\mathcal{R}=\{r_1,\ldots,r_P\}, 
\end{align}
where each positive integer $r_k$ is associated with a subspace  $\mathcal{K}_{\kro_k}(H,\ket{\psi_{k-1}})$ and 
$\ket{\psi_{k}}$ denotes the ground state 
in $\mathcal{K}_{\kro_k}(H,\ket{\psi_{k-1}})$. 
The iterative construction implies that for all $\kro_k\in \mathcal{R}$, we have 
\begin{align}
 \mathcal{K}_{\kro_k}(H,\ket{\psi_{k-1}}) \subseteq \mathcal{K}_{O(\mathcal{R})}(H,\ket{\phi_0})\,,
\end{align}
where we define the order of the partition $\mathcal{R}$ as 
\begin{align}
O(\mathcal{R}) =1-P+\sum_{k=1}^P r_k \,.
\end{align}
Consequently, each QSE problem in $\mathcal{K}_{\kro_k}(H,\ket{\psi_{k-1}})$ can be solved with the matrix elements in Eq.~(\ref{eq:qse_mat}) 
required for building the subspace $\mathcal{K}_{R}(H,\ket{\phi_0})$. However, the ground state in 
$\mathcal{K}_{R}(H,\ket{\phi_0})$ will   in general 
be different from the final ground state $\ket{\psi_P}$ obtained by PQSE. Similarly, the ground state energies of the two approaches 
will be different as well.  Without noise and with infinite numerical precision, the direct solution obtained by 
solving the QSE problem in $\mathcal{K}_{R}(H,\ket{\phi_0})$ is expected to be better. However, in the presence of noise and finite numerical 
precision this need not be the case. The advantage of PQSE is that one has to solve a sequence of smaller, better conditioned, QSE problems.

The idea of our partitioned QSE algorithm is to devise a systematic way of finding $P$ and a sequence $\mathcal{R}$ with  $O(\mathcal{R}) \leq R$ such that  the ground state energy estimate $E_P$ associated with $\ket{\psi_P}$ is a better approximation to the numerical 
estimate of the ground state energy in $\mathcal{K}_{\kroF}(H,\ket{\phi_0})$.  
As explained in the next section, we propose to achieve this by using the variance in Eq.~(\ref{eq:var}) 
to gauge the quality of the solution of each small QSE problem.

\subsubsection{PQSE Algorithm definition} \label{sec:algo_intro}
This algorithm returns an estimate $E_g$ of the ground state energy of a Hamiltonian $H$, as well as a description of the approximate ground state. 
For a given Krylov subspace order $\kroF$ it finds a partition into smaller 
QSE problems $\mathcal{R}$ [see Eq.~(\ref{eq:pqse_seq})] such that all 
small Krylov subspaces are part of $\mathcal{K}_{R}(H, \ket{\phi_0})$.
The PQSE algorithm is as follows.
\begin{enumerate}
    \item Input: Hamiltonian $H$,  Krylov subspace of order $\kroF$ with  initial reference state $\ket{\phi_0}$. The matrix elements defined in Eq.~(\ref{eq:qse_mat}) are obtained using a 
    quantum computer.
    \item Set $\kroM=\kroF$, $P=0$,  $\ket{\psi_0}=\ket{\phi_0}$, 
    $E_g=\bra{\psi_0}H\ket{\psi_0}$. 
    \item While ($1-P+\sum_{k=1}^P r_k < \kroF)$ do:\\
    \begin{enumerate}
    \item Replace $P$ with $P+1$.
    \item For all integers $q$ with  $1 \le q \le \kroM$ do: 
    \begin{enumerate}
    \item Create the generalised eigenvalue problem 
    in $\mathcal{K}_{q}(H,\ket{\psi_{P-1}})$ [see Appendix~\ref{app:recon_ham}].
    \item Solve the generalised eigenvalue problem in $\mathcal{K}_{q}(H, \ket{\psi_{P-1}})$, giving coefficients $\mathbf{c}_q[\ket{\psi_{P-1}}]$ of 
    the ground state estimate $\ket{q}$ relative to the Krylov basis starting from $\ket{\psi_{P-1}}$. 
    \item Calculate the variance  $\text{var}(H,\ket{q})$ [see Appendix~\ref{app:recon}].
    \end{enumerate}
     \item Set $\kro_P = \text{argmin}_{1\le q \le \kroM}\text{var}(H,\ket{q})$.
    \item Set $\ket{\psi_P} = \ket{\kro_P}$ as the ground state of the QSE problem in  $\mathcal{K}_{\kro_P}(H, \ket{\psi_{P-1}})$.

    \item if [$\text{var}(H,\ket{\psi_{P-1}})<\text{var}(H,\ket{\psi_P})$]: \\
    \hspace*{1cm} break \\
    else update 
        \begin{enumerate}
            \item[$\bullet$] $E_g=\bra{\psi_P}H\ket{\psi_P}/\braket{\psi_P\vert\psi_P}$.
            \item[$\bullet$] $\kroM$ with $\kroM-(r_P-1)$.
        \end{enumerate}
    \end{enumerate}
    \item Return $E_g$, $\ket{\psi_P}$
\end{enumerate}

We now quantify the resources used in PQSE. 
PQSE  uses all matrix elements 
of the full Krylov subspace of 
order $R$. 
In addition, calculating the variance may require the additional matrix element $\bra{\phi_0} H^{2 \kroF} \ket{\phi_0}$.

All QSE techniques incur the classical cost of
constructing and solving 
Eq.~(\ref{eq:gev}). The additional classical cost of PQSE comes 
from combining the output of 
previous QSE problems with quantum 
expectation values to produce new 
QSE problems. While this cost may vary based on noise levels and problem parameters, 
we find that the classical cost scales as $\mathcal{O} (R^4)$, see Appendix~\ref{app:cost}. QSE and TQSE incur an $\mathcal{O} (R^3)$ classical cost to solve the generalised eigenvalue problem of Eq.~\eqref{eq:gev} \cite{parlettSymmetricEigenvalueProblem1998}.

The generalisation of the algorithm description to a Krylov basis generated by an operator $A~\neq~H$ is straightforward, see Appendix \ref{app:recon}.

\subsection{Noise model}
We now describe how we model noise on each matrix element in Eq.
(\ref{eq:qse_mat}). Noisy versions of $\mathbf{H}$ and $\mathbf{S}$ are constructed as
\begin{subequations}
\label{eq:qse_mat_noise}
\begin{align}
\tilde{\mathbf{H}} & = \mathbf{H} + \mathbf{\Delta}_H\,, \\
\tilde{\mathbf{S}} & = \mathbf{S} + \mathbf{\Delta}_S\,,
\end{align}
\end{subequations}
where $\mathbf{\Delta}_H$ and $\mathbf{\Delta}_S$ are Hankel matrices with elements 
$\Delta_{H, ij}$ and $\Delta_{S, ij}$ which quantify the uncertainty in $\bra{\phi_0} H^{i + j + 1}\ket{\phi_0}$ and $\bra{\phi_0} H^{i + j}\ket{\phi_0}$.
Because $\mathbf{H}$ contains all elements of $\mathbf{S}$ except $S_{00} = \braket{\phi_0 \lvert \phi_0}=1 $, 

we only consider constructing $\mathbf{\Delta}_{H}$.
We draw $\mathbf{\Delta}_{H, ij}$ from a normal distribution of width $\sigma_{H, ij}$ where
\begin{align}
    \sigma_{H, ij} = \delta \sqrt{\bra{\phi_0} H^{2(i+j+1)}\ket{\phi_0} - \bra{\phi_0} H^{i+j+1} \ket{\phi_0}^2} \,,
\end{align}
for a dimensionless parameter $\delta$ which controls noise strength.
By considering measuring each matrix element $1 / \delta^2$ times we see that this choice of 
noise model is equivalent to assuming we have access to unbiased estimators of $H_{ij}$, with any error resulting from finite sampling 
or limited numerical precision. This assumption can be motivated by considering applying this technique in either i) the regime where quantum error 
mitigation can still be effectively applied, but classical simulation is not possible~\cite{kimEvidenceUtilityQuantum2023}; ii) the regime of fault 
tolerant quantum computing. In \cite{kirbyExactEfficientLanczos2023} 
it is suggested that Krylov basis QSE may be a favourable choice over Phase Estimation in the early fault-tolerant regime, highlighting the former's noise robustness as motivation. 
When implementing TQSE to mitigate the effects of noise, we follow~\cite{epperlyTheoryQuantumSubspace2022} and set our thresholding  
parameter $\tau$ to $10^{-a} \left( ( \eta^2_H + \eta^2_S )^{1/2} \right)$ where $\eta_H, \eta_S$ are the spectral 
norms of the perturbations $\mathbf{\Delta}_H, \mathbf{\Delta}_S$ and $a$ is a threshold scaling hyper-parameter we tune to approximate the optimal value of $\tau$.
In order to tune this hyper-parameter we run each TQSE experiment with a given $\delta$ for 50 evenly spaced values of $-0.5 \leq a \leq 5$ and take the value which gives the best solution according to the particular error metric we are using.
This approach uses the noise-free QSE matrices to calculate $\eta_H$, $\eta_S$ and the true ground state energy to minimize $a$ with respect to our error metric.
We highlight that finding the optimal threshold in this way would not be possible without knowledge of the exact solution.

\section{Results} \label{sec:num} 
\begin{figure*}
    \centering
    \includegraphics[width=0.89\textwidth]{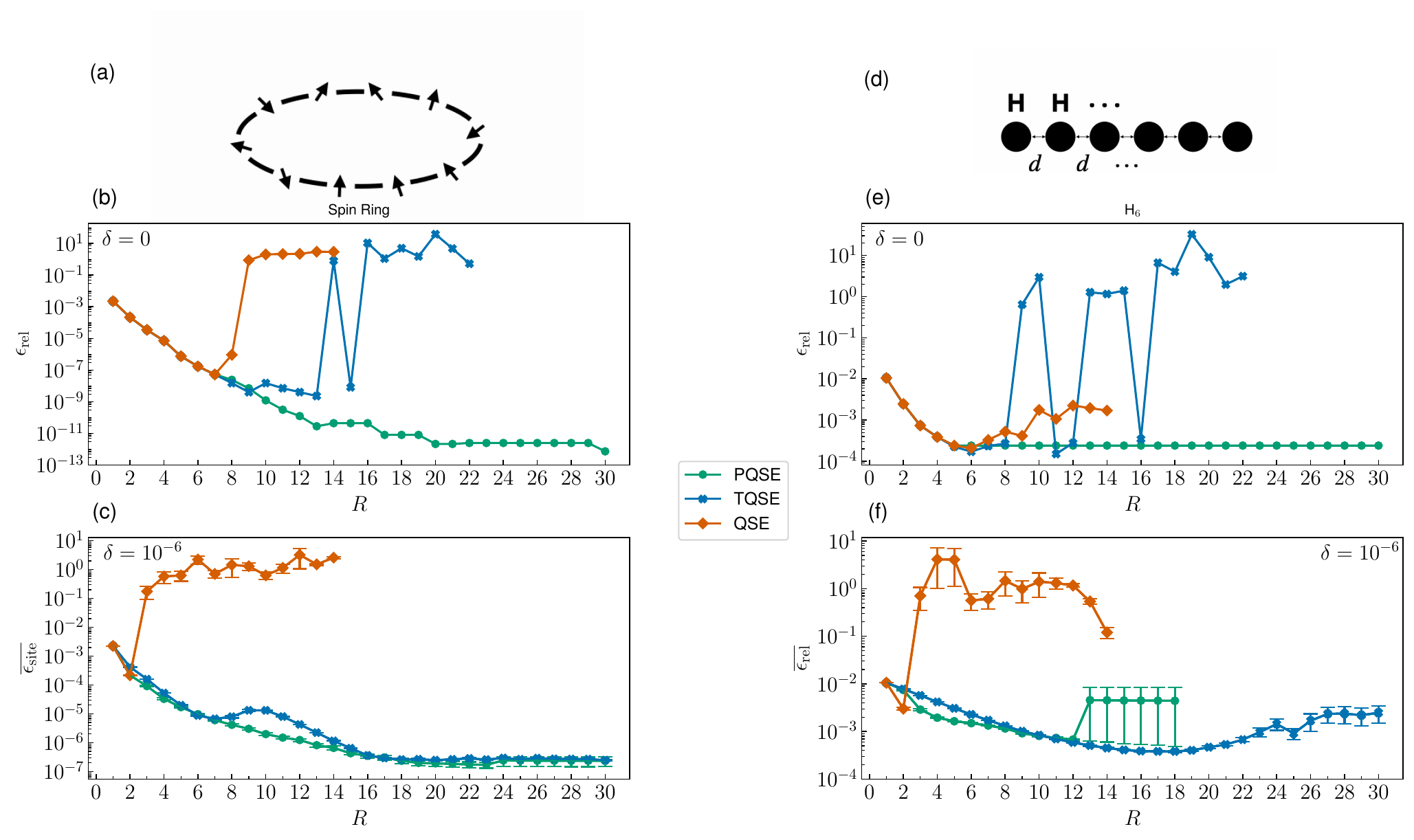}
    \caption{Average relative error in estimate of ground state energy for QSE with increasing Krylov basis of order $R$ generated using Hamiltonian powers: unmodified (QSE), thresholded (TQSE), partitioned (PQSE). Left column: 10 qubit periodic disordered Heisenberg model with $J$ = 0.1, $h$ = 1, the model is depicted in (a) with results for Gaussian noise strength $\delta = 0$ in (b) and $\delta = 10^{-6}$ in (c). Right column: H$_6$ in a STO-3G basis with a parity mapping applied in equilibrium configuration, the system is depicted in (d) where H and $d$ represent Hydrogen molecules and bond length respectively, with results for Gaussian noise strength $\delta = 0$ in (e) and $\delta = 10^{-6}$ in (f).
    } 
    \label{fig:err_decay}
\end{figure*}

We now provide two example problems to enable a comparison between the different QSE techniques we have presented. 
The first example is the $n$ qubit periodic Heisenberg model with a random magnetic field in the $z$ direction, 
see Fig.~\ref{fig:err_decay}(a) for a schematic and Fig.~\ref{fig:err_decay}(b, c) for results. This is modelled by the Hamiltonian
\begin{equation} \label{eqn:1d_spin_ring_disord}
    H_{\text{ring}} = \sum_{i}^{n} J \boldsymbol{\sigma}_i \cdot \boldsymbol{\sigma}_{i+1}  + h_i \sigma^z_i,
\end{equation}
where $ \boldsymbol{\sigma}_i = (\sigma^x_i, \sigma^y_i, \sigma^z_i)$ is a vector of Pauli matrices acting on qubit $i$ and $\boldsymbol{\sigma}_{n+1} 
=  \boldsymbol{\sigma}_{1}$. $h_i$ are uniformly drawn from $(-h,h)$ where $h$ quantifies the magnitude of disorder present in the system. 
We measure energy in units of $h$ and  set $h = 1$ and $n=10$.  In the regime  $h \gg J$, many body localization has been argued to emerge~\cite{nandkishoreManyBodyLocalizationThermalization2015}. 
For fixed values $h_i$ we construct $\ket{\phi_0}$ by solving Eq.(\ref{eqn:1d_spin_ring_disord}) with $J = 0$, a simplification which makes the 
model efficiently solvable on a classical computer. The solution can be prepared on a quantum computer through a single layer 
of bit flip gates. 
The problem of finding the ground state of this model has been proposed as a candidate for displaying a 
quantum advantage which is more accessible to less advanced quantum computers \cite{childsFirstQuantumSimulation2018}. It is a 
well studied model in the context of self-thermalization and many body localization in condensed matter physics. The difficulty 
of classically approximating the ground state has placed an obstacle to further understanding as $n$ grows 
\cite{luitzManybodyLocalizationEdge2015}.

The second example system is given by an electronic structure Hamiltonian for the $\text{H}_6$ chain, see Fig.~\ref{fig:err_decay}(d) for a schematic and Fig.~\ref{fig:err_decay}(e, f) for results. The electronic structure Hamiltonian for H$_6$ can be written as
\begin{align}
H_{\text{H}_{6}}=\sum_{i, j} h_{i j} a_i^{\dagger} a_j+\frac{1}{2} \sum_{i, j, k, l} h_{i j k l} a_i^{\dagger} a_j^{\dagger} a_k a_l\,.
\end{align}
The fermionic creation and annihilation operators $a_i^{\dagger}$ and $a_i$ act on the $i$th spin orbital and obey $\{a_i, a_j\} = 0$, $\{a_i^{\dagger}, a_j^{\dagger}\} = 0$ and $\{a_i, a_j^{\dagger}\} = \delta_{ij} \mathbf{I}$  where $\{A, B\} = AB + BA$.  $h_{ij}$ and $h_{ijkl}$ are one and two electron overlap integrals respectively. We construct the molecule in the STO-3G basis~\cite{szaboModernQuantumChemistry1996}, in equilibrium configuration at a bond length of 1.5 ~\r{A}, and use a parity mapping~\cite{seeleyBravyiKitaevTransformationQuantum2012} to convert from a fermionic to a 
10 qubit Hamiltonian. We construct $\ket{\phi_0}$ as the Hartree-Fock state. Molecular properties were computed using the PySCF library~\cite{sun2018pyscf}. The strong electron correlation which arises in this system can prove challenging for classical methods to simulate as additional Hydrogen atoms are considered~\cite{stairMultireferenceQuantumKrylov2020}. 

We now present the three different error metrics we will use.
For comparisons of the different QSE techniques in the absence of noise ($\delta=0$), we vary the 
Krylov basis order $R$ and compute the relative error in the ground state energy
\begin{align}
\epsilon_{\text{rel}} = \left|\frac{E_g - \tilde{E}_g}{\tilde{E}_g}\right|\,,
    \label{eq:relerr}
\end{align}
where $\tilde{E}_g$ is the ground state energy obtained by exact diagonalization and $E_g$ is obtained by either QSE, TQSE or PQSE.  
For comparisons at a non-zero noise strength $\delta$
our performance metric is 
denoted by $\overline{\epsilon_{\text{rel}}}$,
where the overbar denotes the average of  $\epsilon_{\text{rel}}$ in Eq.~(\ref{eq:relerr}) over 200 noise instances.
The associated uncertainty in $\overline{\epsilon_{\text{rel}}}$ is the standard error. 
In order to investigate the performance of each approach to QSE as a function of noise strength, we compute 
$\overline{\epsilon_{\text{rel}}}$ for a fixed noise strength, again over 200 noise instances, and find its minimum value as a function  of $R$,
\begin{align}
\aveErr =\min\limits_R \overline{\epsilon_{\text{rel}}}(R)\,.
\end{align}
Without noise and infinite numerical precision this minimum would be lower bounded by $\xi=0$, as one could increasing the subspace 
dimension indefinitely, subject to the available quantum resources. However, with noise and finite numerical precision, there will be a limit to the size of Krylov subspace dimension that can be reliably used. Therefore, this minimum provides an indicator of the lowest possible average error achieved by the considered QSE techniques.

\textit{Heisenberg model.} The results for $\epsilon_{\text{rel}}$ as a function of $R$  in case of the spin ring are shown in 
Fig.~\ref{fig:err_decay}(b). We find that the relative error initially decreases exponentially with $R$ for all three techniques. 
In spite of the absence of intentionally added noise, finite machine precision introduces a barrier to arbitrarily low solution error 
and results in numerical instability. In order to mitigate the effect of noise due to finite machine precision we fix the TQSE threshold parameter to $\tau = 10^{-13}$, a value found to be effective in other work~\cite{kirbyExactEfficientLanczos2023}. Note that thresholding is not used as part of PQSE.
With this, TQSE allows one to explore larger values of $R$ 
and the relative error is approximately one order of magnitude smaller compared to QSE. 
PQSE can have a slighter higher average error at basis orders where QSE fails and TQSE is numerically stable. On the other hand, PQSE returns 
more accurate solutions in regions where the other techniques fail. We find that by increasing $R$, PQSE can lower $\epsilon_{\text{rel}}$ 
by three orders of magnitude compared to TQSE. 

We now move on to results with a fixed noise strength $\delta = 10^{-6}$, shown in Fig.~\ref{fig:err_decay}(c).
In contrast to the $\delta = 0$ example, we see that numerical instability restricts use of QSE to Krylov basis orders of 1 or 2. Implementing thresholding allows an exponentially fast decrease in average error for a limited range of $R$. For $8 \leq R \leq 11$ we see that the average error of TQSE increases before reconverging. The average error of PQSE decreases to a minimum at $R=23$ which is marginally lower than that of TQSE.
We provide insight into the subspace dimensions being discarded by TQSE, as well as the typical sequences chosen by PQSE in this experiment in Appendix~\ref{app:dim}.

\begin{figure*}
    \centering
    \includegraphics[width=\textwidth]{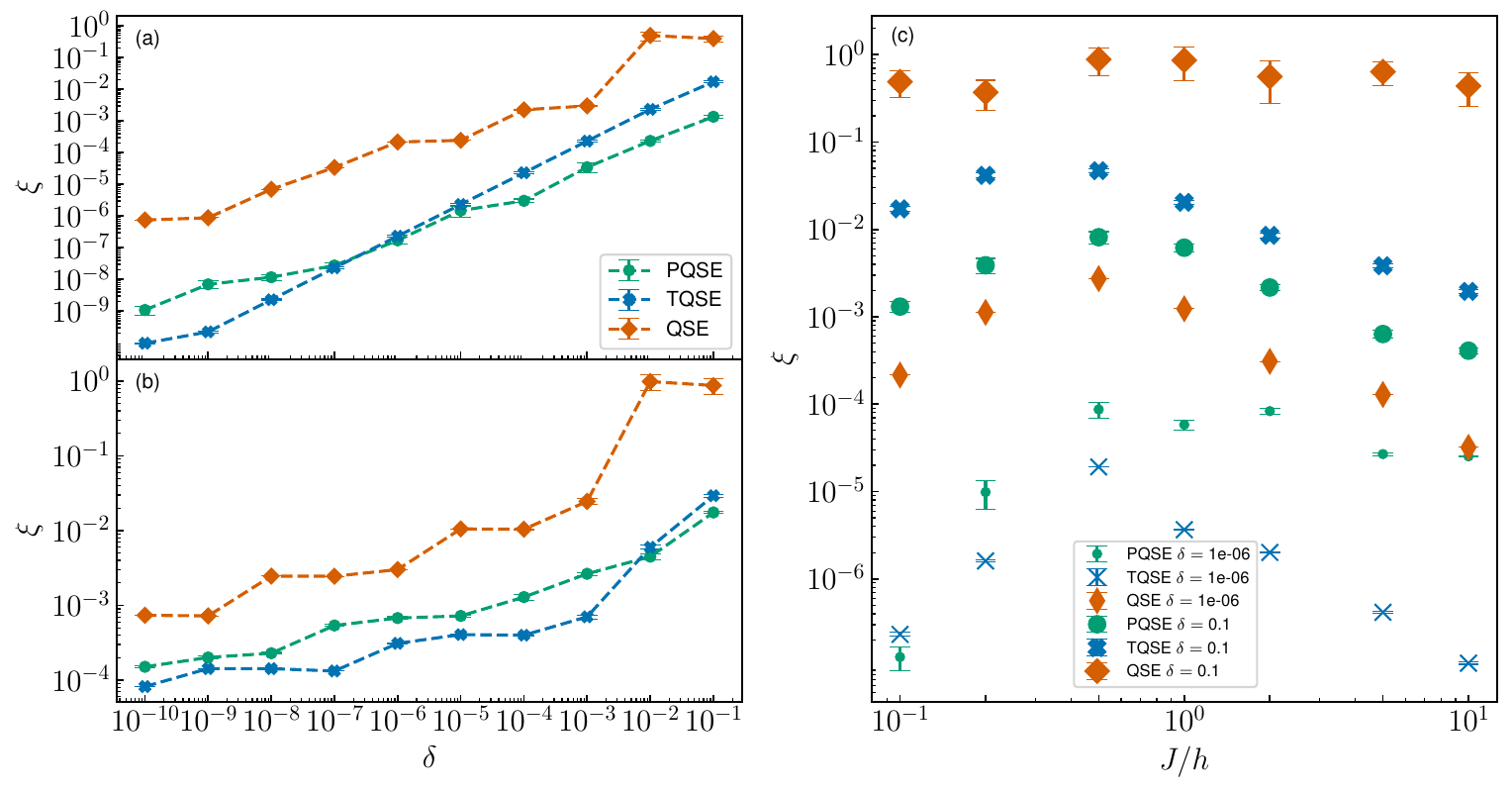}
    \caption{
    Minimum average relative error $\xi$ in the estimate of ground state energy  for QSE with a Krylov basis generated using Hamiltonian powers: unmodified (QSE), thresholded (TQSE), partitioned (PQSE). 
    (a) $\xi$ as a function of noise strength $\delta$ for a  10 qubit periodic disordered Heisenberg model with $J/h=0.1$.  (b)  $\xi$ as a function of noise strength $\delta$ for H$_6$ in a STO-3G basis with a parity mapping applied in equilibrium configuration. (c)  $\xi$ as a function of $J/h$ for the 10 qubit periodic disordered Heisenberg model and noise strengths $\delta=10^{-6},10^{-1}$. Error bars show the standard error.}
   \label{fig:min_ave_error}
\end{figure*}

The results for $\aveErr$ as a function of noise strength are shown in Figs.~\ref{fig:min_ave_error}(a). 
We find that PQSE performs competitively with TQSE across the range of noise strengths we consider, with $\aveErr$ between half an order and one and a half orders of magnitude lower for TQSE than PQSE for $10^{-10} \leq \delta \leq 10^{-8}$. The two approaches perform roughly the same at $10^{-7} \leq \delta \leq 10^{-5}$. At $10^{-4} \leq \delta \leq 10^{-1}$, $\aveErr$ for PQSE is approximately an order of magnitude lower than TQSE.

In order to both better understand the spin ring example and the behaviour of PQSE with different choices of reference state, we apply each QSE algorithm to solving Eq.~(\ref{eqn:1d_spin_ring_disord}) for a range of values of $J/h$. We use the $J=0$ solution as an initial state $\ket{\phi_0}$ for $J/h = 0.1, 0.2$ and the $h=0$ solution for larger ratios.
The results for $\aveErr$ are shown in Fig.~\ref{fig:min_ave_error}(c) where, for clarity, we include two noise strengths $\delta = 10^{-6}, 10^{-1}$.
We see that at  $\delta = 10^{-1}$, $\aveErr$ for PQSE is between half an order to an order of magnitude lower than TQSE. At $\delta = 10^{-6}$, $\aveErr$ for TQSE is lower than that for PQSE for $J/h > 10^{-1}$ from half an order to two orders of magnitude.
All techniques display greater variation in $\aveErr$ across the range of $J/h$ for weaker noise compared with the values of $\aveErr$ for stronger noise. This suggests that for weaker noise the numerical stability of each QSE technique, particularly PQSE, is more sensitive to the properties of $H$ and the choice of initial state $\ket{\phi_0}$, than for stronger noise.

\textit{Hydrogen chain.}
The results for $\epsilon_{\text{rel}}$ as a function of $R$  in case of the $\text{H}_6$ chain are shown in 
Fig.~\ref{fig:err_decay}(e). 
We observe that $\epsilon_{\text{rel}}$ decays much slower with $R$ as compared to 
the spin ring for all QSE techniques. This finding is consistent with the results in~\cite{kirbyExactEfficientLanczos2023}, where 
the slow convergence with $R$ was explained by the use of a Hartree-Fock state as a reference state. 
QSE and TQSE display similar qualitative behaviour to the spin ring example. In contrast to the spin ring example, PQSE does not display signs of numerical instability, instead converging to a fixed value of $\epsilon_{\text{rel}}$. 
This fixed value is higher than the minimum error achieved by QSE and TQSE. 
We obtain insight into this by computing the overlap of each QSE solution with the true ground state found using exact diagonalisation. At $R=6$, QSE returns a lower energy error than PQSE, but the 
QSE ground state estimate has a smaller overlap with the true ground state than the PQSE estimate due to numerical instabilities leading to an inconsistency between $E$ and $\bra{\psi} H \ket{\psi}$ after solving Eq.~(\ref{eq:gev}). On the other hand, the overlap of the ground state estimate obtained 
by TQSE for $R=6$ and $R=7$ is larger than the corresponding values for PQSE, which is consistent with the better energy estimates of TQSE at these points. 

Results for a fixed noise strength $\delta = 10^{-6}$ are shown in Fig.~\ref{fig:err_decay}(f).
We see a monotonic decrease in TQSE error to a minimum at $\overline{\epsilon_{\text{rel}}} = 2 \times 10^{-4}$, followed by a gradual increase. Here PQSE achieves a minimum error of approximately $\overline{\epsilon_{\text{rel}}} = 7 \times 10^{-4}$.
The increase in $\overline{\epsilon_{\text{rel}}}$ and its variance for PQSE at $\kroF=13$ is due to an ill-conditioned eigenvalue problem.
As in the noiseless case, we find that QSE methods are more effective for the spin ring example than for $\text{H}_6$.
The results for $\aveErr$ as a function of noise strength are shown in Figs.~\ref{fig:min_ave_error}(b).
PQSE slightly outperforms TQSE for $10^{-2} \leq \delta \leq 10^{-1}$ and TQSE outperforms PQSE for $10^{-10} \leq \delta \leq 10^{-3}$ by at most approximately half an order of magnitude. 

\begin{figure*}[!t]
  \centering
      \includegraphics[width=\textwidth]{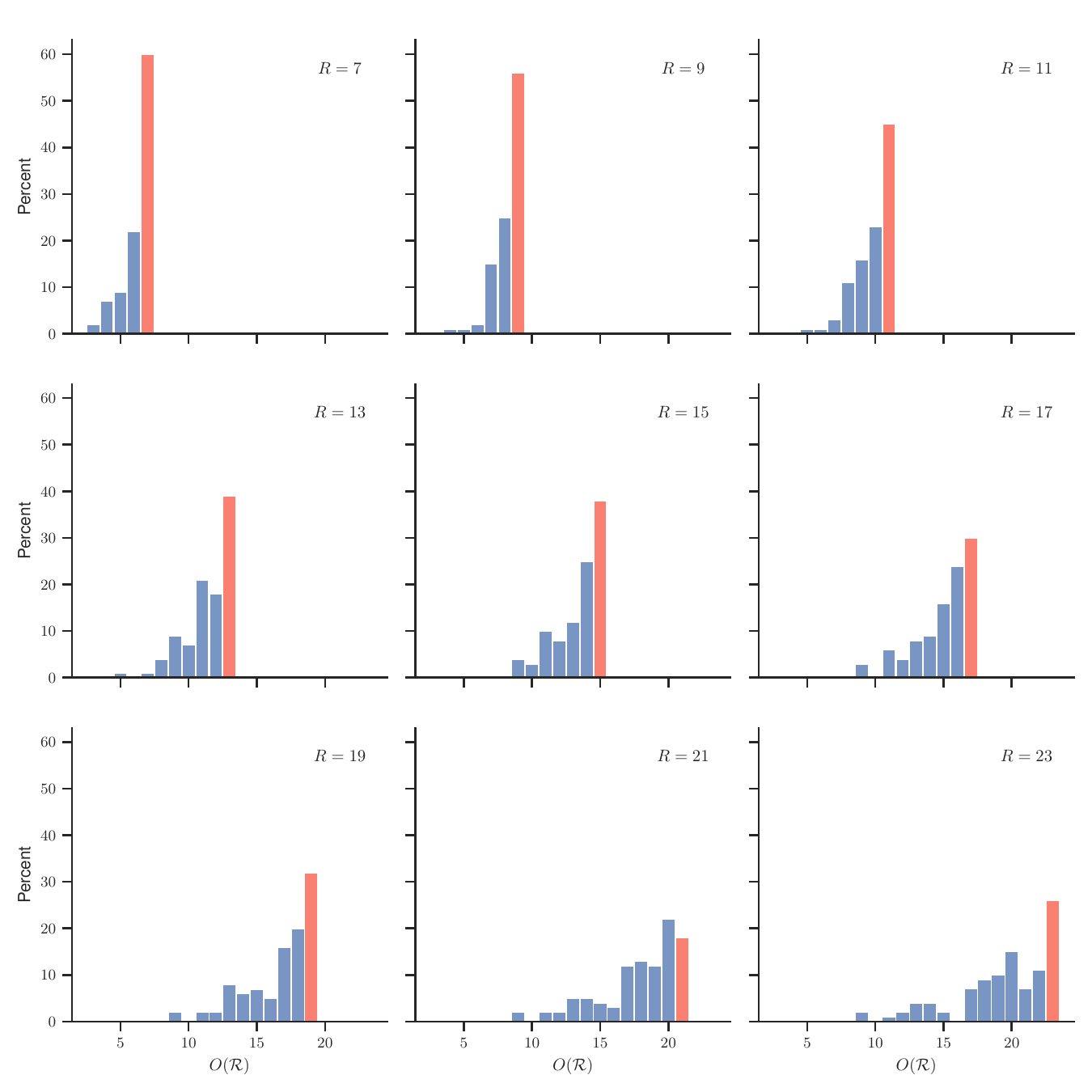}
  \caption{Distribution of the Krylov basis order of the final basis returned by PQSE $O(\mathcal{R})$ for increasing maximum allowed Krylov basis order $R$. Data for each distribution is obtained over 200 instances of a noisy 10 qubit periodic disordered Heisenberg model. Noise is drawn from Gaussian distributions width proportional to $\delta = 10^{-6}$. The maximum Krylov basis order is the far right bin, coloured in red.}
  \label{fig:kbo_termination}
\end{figure*}

We now present a brief investigation into the typical behaviour of PQSE.
As noted in Sec.~\ref{sec:algo_intro}, the order $O(\mathcal{R})$ of 
the partition $\mathcal{R}$ found by PQSE may be smaller than  
the maximally allowed  Krylov basis order  $\kroF$. 
We now quantify how frequently $O(\mathcal{R})$ is smaller than $R$ and analyse the 
pattern of values of $O(\mathcal{R})$. Figure~\ref{fig:kbo_termination} shows the distribution of $O(\mathcal{R})$ for increasing $\kroF$ 
for the spin ring example at fixed noise strength.  The figure illustrates the relationship between $O(\mathcal{R})$ and $\kroF$ as the 
latter increases. We see that the proportion of instances displaying early termination increases as $\kroF$ increases and 
that the range of basis orders at early termination also increases. 
As the target basis 
order for PQSE increases, a large proportion of instances terminate with a basis order above this minimum. Figures \ref{fig:err_decay} and \ref{fig:min_ave_error} show that PQSE can continue increasing accuracy at basis orders above this minimum. This suggests that this implementation of PQSE is capable of
detecting and avoiding numerical instabilities by choosing a stable basis sequence or terminating early. 
An extension to these results can be found in Appendix~\ref{app:dim}, where we provide overview of the typical basis-size sequences chosen by PQSE for the spin ring example with $\delta = 10^{-6}$.

Additional results for $\overline{\epsilon_{\text{rel}}}$ as a function of $R$ for a range of noise strengths are provided in Appendix~\ref{app:add_num} and show similar qualitative behaviour to the $\delta=10^{-6}$ case.
Furthermore, in Appendix~\ref{app:add_num} we also recompute Figs.~\ref{fig:min_ave_error}(a) and (b) for the Heisenberg model and $\text{H}_6$ examples using a PQSE criterion which avoids using the matrix element $\bra{\phi_0} H^{2 \kroF} \ket{\phi_0}$, finding performance to be very similar in both cases.
Finally, in order to present results in a more near term relevant setting, as well as providing evidence that PQSE generalize to other Krylov basis generation methods, we present the results of our numerical experiments with a Krylov basis generated through real time evolution in Appendix~(\ref{app:rte}). We see comparative performance between PQSE and parameter-optimized TQSE, motivating application of PQSE in near-term experiments.

\section{Discussion} \label{sec:disc} 
We have presented a classical post-processing technique that
allows the QSE algorithm to be effectively utilised with a higher dimension Krylov subspace without requiring problem and noise dependent parameter tuning.
The practical implication of this is that it can enable
more accurate ground state energy estimates.
In addition to competitive performance with the state of the art, PQSE can offer order of magnitude improvements in accuracy.

The two presented  examples probed the accuracy of each  QSE variant in the presence of finite sampling error. 
In particular, we applied all three QSE approaches 
to estimating the ground state energy of a quasi-random spin system and a molecular Hamiltonian. 
We found that for the spin ring example at smaller ratios of $J/h$ and the molecular example PQSE either performed better or comparatively to TQSE with a optimized parameterisation. As the TQSE threshold was chosen by optimising with respect to the true ground state energy, we argue that this is close to the best-case performance for TQSE. That PQSE remains competitive under these conditions is particularly noteworthy.
For larger ratios of  $J/h$, TQSE was more accurate than PQSE.
Obtaining a clearer understanding of this discrepancy, perhaps whether it is due to sensitivity to the choice of $\ket{\phi_0}$, is an interesting starting point for future work.
The range of minimum average errors achieved by all techniques in the spin ring example with $J/h=0.1$ and the molecular system were substantially different. 
This highlights that, while PQSE is able to construct sequences of eigenvalue problems with better conditioning, it is unable to overcome limitations in accuracy set by a poor choice of initial state.
Improvements to the initial state, or techniques to augment the standard Krylov basis [see Eq.~(\ref{eq:krylov_basis})] with improved convergence will therefore be highly valuable. 
Additionally, for both PQSE and TQSE, we find that the variational principle constraining the trial solution energy to be less than the true ground state energy of our target Hamiltonian is broken. Ensuring this principle holds would be a useful improvement for future work.
The iterative construction we have proposed leaves considerable freedom to implement these improvements. For example, one could vary the criteria used to evaluate good sequences throughout the algorithm, perhaps by initially rewarding exploration before transitioning to a more targeted criterion such as the variance. Alternatively, one could consider constructing composite subspace sequences, employing the Krylov subspace when most effective. 
These improvements could draw from classical iterative Krylov approaches such as the Restarted Arnoldi algorithm \cite{lehoucq1996deflation}, which also uses reference state re-initialisation, and the Davidson algorithm \cite{DAVIDSON197587}, which has already seen development in a quantum context \cite{tkachenkoQuantumDavidsonAlgorithm2024, berthusenMultireferenceQuantumDavidson2024} and involves carefully growing a Krylov basis through a process of computing residual vectors and applying a pre-conditioner.

In order to place our new approach in context we  briefly discuss PQSE in relation to an existing iterative approach to QSE, the Iterative Quantum Assisted Eigensolver (IQAE)~\cite{bhartiIterativeQuantumassistedEigensolver2021}. IQAE proceeds by using the $M$ term unitary decomposition, $H = \sum_{i=1}^M c_i U_i$, to decompose powers of the target Hamiltonian $H$ as 
\begin{align}
H^k \ket{\phi_0} & = \left(\sum_{i=1}^M c_i U_i\right)^k \ket{\phi_0}\\
& =\sum_{\{i_j\}_{j=1}^{k} = 1}^M c_{i_1} \dots c_{i_k} U_{i_1}\dots U_{i_k} \ket{\phi_0}.
\end{align}
This decomposition is used to form a basis 
as $\mathcal{B}_k = \{ \ket{\phi_0}\} \cup \{ U_{i} \ket{\phi_0} \}_{i=1}^M \cup \dots \cup \{ U_{i_1} U_{i_2} \dots U_{i_k} 
 \ket{\phi_0} \}_{i_1, i_2, \dots, i_k = 1}^M$, to be contrasted with Eq.~(\ref{eq:krylov_basis}). Using $\mathcal{B}_k$ as a basis for QSE avoids the need to directly embed $H$ during a quantum computation. 
 The lowest energy weighting of elements in $\mathcal{B}_k$ is computed at increasing 
$k \geq 0$ until a desired convergence condition is met. 

Both IQAE and PQSE are iterative subspace methods, but the respective subspaces are constructed 
differently. Our approach is motivated by the work 
in~\cite{kirbyExactEfficientLanczos2023} which presents an efficient algorithm to construct an effectively exact 
Krylov subspace. This is to be contrasted with the unitary decomposition used by IQAE. Furthermore, whereas IQAE works with a fixed reference state at each iteration, our procedure uses 
the solution from a previous iteration as a new effective reference state for the next. 

A special feature of IQAE is that the basis $\mathcal{B}_k$ is constructed 
from unitaries which do not conserve the symmetry of $H$. Therefore, if the initial state 
$\ket{\phi_0}$ lies in a different symmetry sector to the ground state of $H$, 
then optimising using $\mathcal{B}_k$ can lower the energy of $\ket{\phi_0}$ while a Krylov basis constructed with 
a symmetry-preserving Hamiltonian $H$ would fail. An interesting avenue for future research is 
to consider applying PQSE with the basis used with the IQAE algorithm, to see if the dual benefits of 
numerical stability and symmetry breaking can be combined. 
A similar direction could be using classical shadow data, which can be applied to enhance the performance of subspace methods with a poor initial state \cite{boydHighdimensionalSubspaceExpansion2025}.
Further improvements of PQSE could 
include augmenting the variance criterion for choosing a good sequence of partitions, 
and by investigating how PQSE may be more tightly integrated with thresholding. 

A strategy for applying PQSE to a system where exact diagonalisation is not possible would be to gradually increase the Krylov basis size  until convergence is 
reached, or the markers of numerical instability appear. In particular, 
it is beneficial to apply PQSE together with standard QSE and TQSE methods 
in order to increase confidence in the obtained results. Since  
PQSE can be implemented using the same matrix elements required for QSE and TQSE, with an additional element required if the proposed variance minimisation criterion is used, this can be done with a small increase in quantum resources. 
Insight into the ability of PQSE to operate at classically intractable problem sizes can be found in a recent study of the application of QSE to solving Lattice Gauge Theories \cite{andersonSolvingLatticeGauge2025}.

Recent experimental demonstrations of quantum error correction~\cite{bluvsteinLogicalQuantumProcessor2023, acharyaSuppressingQuantumErrors2023, krinnerRealizingRepeatedQuantum2022a, ryan-andersonImplementingFaulttolerantEntangling2022a} provide some of the strongest evidence so far that the technological obstacles to fault tolerant quantum computation can be overcome. However, even the accuracy of early fault tolerant quantum computers may be limited by the impact of finite numerical precision and statistical noise on the output of algorithms such as QSE. 
Our approach suggests that this can be strongly alleviated in a parameter-free way.

\section{Code Availability}
Code for our implementation of PQSE
can be found at \url{https://github.com/t1491625/pqse_demo}, where Fig.~\ref{fig:err_decay}(c) is reproduced as a guiding example.

\section{Acknowledgements}
 T. O. acknowledges support by the EPSRC through an EPSRC iCASE studentship award in collaboration with IBM Research (EP/W522211/1) and the Laboratory Directed Research and 
 Development (LDRD) program of Los Alamos National Laboratory (LANL) under project number 20230049DR.
 D. J. acknowledges support by the European Union’s Horizon Programme (HORIZON-CL4-2021-DIGITALEMERGING-02-10) Grant Agreement 101080085 QCFD. 
  D. J. acknowledges support from the Hamburg Quantum Computing Initiative (HQIC) project EFRE. The project is co-financed by ERDF of the European Union and by “Fonds of the Hamburg Ministry of Science, Research, Equalities and Districts (BWFGB)
 The authors would 
 like to acknowledge the use of the University of Oxford Advanced Research Computing (ARC) facility in carrying out this 
 work~\cite{richardsUniversityOxfordAdvanced2015}.

\bibliographystyle{quantum}
\bibliography{q_lib, lewis_bib}

\begin{thebibliography}{10}

\bibitem{lubaschVariationalQuantumAlgorithms2020}
Michael Lubasch, Jaewoo Joo, Pierre Moinier, Martin Kiffner, and Dieter Jaksch.
\newblock ``Variational quantum algorithms for nonlinear problems''.
\newblock \href{https://dx.doi.org/10.1103/PhysRevA.101.010301}{Physical Review
  A {\bf 101}, 010301}~(2020).
\newblock  \href{http://arxiv.org/abs/1907.09032}{arXiv:1907.09032}.

\bibitem{kyriienkoSolvingNonlinearDifferential2021}
Oleksandr Kyriienko, Annie~E. Paine, and Vincent~E. Elfving.
\newblock ``Solving nonlinear differential equations with differentiable
  quantum circuits''.
\newblock \href{https://dx.doi.org/10.1103/PhysRevA.103.052416}{Physical Review
  A {\bf 103}, 052416}~(2021).
\newblock  \href{http://arxiv.org/abs/2011.10395}{arXiv:2011.10395}.

\bibitem{kreulaNonlinearQuantumclassicalScheme2016}
J.~M. Kreula, S.~R. Clark, and D.~Jaksch.
\newblock ``Non-linear quantum-classical scheme to simulate non-equilibrium
  strongly correlated fermionic many-body dynamics''.
\newblock \href{https://dx.doi.org/10.1038/srep32940}{Scientific Reports {\bf
  6}, 32940}~(2016).
\newblock  \href{http://arxiv.org/abs/1510.05703}{arXiv:1510.05703}.

\bibitem{kreulaFewqubitQuantumclassicalSimulation2016}
Juha~M. Kreula, Laura {Garc{\'i}a-{\'A}lvarez}, Lucas Lamata, Stephen~R. Clark,
  Enrique Solano, and Dieter Jaksch.
\newblock ``Few-qubit quantum-classical simulation of strongly correlated
  lattice fermions''.
\newblock \href{https://dx.doi.org/10.1140/epjqt/s40507-016-0049-1}{EPJ Quantum
  Technology {\bf 3}, 11}~(2016).
\newblock  \href{http://arxiv.org/abs/1606.04839}{arXiv:1606.04839}.

\bibitem{andersonCoarseGrainedIntermolecular2022}
Lewis~W. Anderson, Martin Kiffner, Panagiotis~Kl Barkoutsos, Ivano Tavernelli,
  Jason Crain, and Dieter Jaksch.
\newblock ``Coarse grained intermolecular interactions on quantum processors''.
\newblock \href{https://dx.doi.org/10.1103/PhysRevA.105.062409}{Physical Review
  A {\bf 105}, 062409}~(2022).
\newblock  \href{http://arxiv.org/abs/2110.00968}{arXiv:2110.00968}.

\bibitem{jaderbergRecompilationenhancedSimulationElectron2022}
Benjamin Jaderberg, Alexander Eisfeld, Dieter Jaksch, and Sarah Mostame.
\newblock ``Recompilation-enhanced simulation of electron--phonon dynamics on
  {{IBM}} quantum computers''.
\newblock \href{https://dx.doi.org/10.1088/1367-2630/ac8a69}{New Journal of
  Physics {\bf 24}, 093017}~(2022).

\bibitem{jakschVariationalQuantumAlgorithms2023}
Dieter Jaksch, Peyman Givi, Andrew~J. Daley, and Thomas Rung.
\newblock ``Variational {{Quantum Algorithms}} for {{Computational Fluid
  Dynamics}}''.
\newblock \href{https://dx.doi.org/10.2514/1.J062426}{AIAA Journal {\bf 61},
  1885--1894}~(2023).

\bibitem{mottaSubspaceMethodsElectronic2024}
Mario Motta, William Kirby, Ieva Liepuoniute, Kevin~J Sung, Jeffrey Cohn,
  Antonio Mezzacapo, Katherine Klymko, Nam Nguyen, Nobuyuki Yoshioka, and
  Julia~E Rice.
\newblock ``Subspace methods for electronic structure simulations on quantum
  computers''.
\newblock \href{https://dx.doi.org/10.1088/2516-1075/ad3592}{Electronic
  Structure {\bf 6}, 013001}~(2024).

\bibitem{huangQuantumAdvantageLearning2022}
Hsin-Yuan Huang, Michael Broughton, Jordan Cotler, Sitan Chen, Jerry Li, Masoud
  Mohseni, Hartmut Neven, Ryan Babbush, Richard Kueng, John Preskill, and
  Jarrod~R. McClean.
\newblock ``Quantum advantage in learning from experiments''.
\newblock \href{https://dx.doi.org/10.1126/science.abn7293}{Science {\bf 376},
  1182--1186}~(2022).

\bibitem{cerezoVariationalQuantumFidelity2020}
M.~Cerezo, Alexander Poremba, Lukasz Cincio, and Patrick~J. Coles.
\newblock ``Variational {{Quantum Fidelity Estimation}}''.
\newblock \href{https://dx.doi.org/10.22331/q-2020-03-26-248}{Quantum {\bf 4},
  248}~(2020).
\newblock  \href{http://arxiv.org/abs/1906.09253}{arXiv:1906.09253}.

\bibitem{cerezoVariationalQuantumAlgorithms2021}
M.~Cerezo, Andrew Arrasmith, Ryan Babbush, Simon~C. Benjamin, Suguru Endo,
  Keisuke Fujii, Jarrod~R. McClean, Kosuke Mitarai, Xiao Yuan, Lukasz Cincio,
  and Patrick~J. Coles.
\newblock ``Variational {{Quantum Algorithms}}''.
\newblock \href{https://dx.doi.org/10.1038/s42254-021-00348-9}{Nature Reviews
  Physics {\bf 3}, 625--644}~(2021).
\newblock  \href{http://arxiv.org/abs/2012.09265}{arXiv:2012.09265}.

\bibitem{peruzzoVariationalEigenvalueSolver2014}
Alberto Peruzzo, Jarrod McClean, Peter Shadbolt, Man-Hong Yung, Xiao-Qi Zhou,
  Peter~J. Love, Al{\'a}n {Aspuru-Guzik}, and Jeremy~L. O'Brien.
\newblock ``A variational eigenvalue solver on a photonic quantum processor''.
\newblock \href{https://dx.doi.org/10.1038/ncomms5213}{Nature Communications
  {\bf 5}, 4213}~(2014).

\bibitem{stairMultireferenceQuantumKrylov2020}
Nicholas~H. Stair, Renke Huang, and Francesco~A. Evangelista.
\newblock ``A {{Multireference Quantum Krylov Algorithm}} for {{Strongly
  Correlated Electrons}}''.
\newblock \href{https://dx.doi.org/10.1021/acs.jctc.9b01125}{Journal of
  Chemical Theory and Computation {\bf 16}, 2236--2245}~(2020).

\bibitem{mccleanHybridQuantumClassicalHierarchy2017}
Jarrod~R. McClean, Mollie~E. Schwartz, Jonathan Carter, and Wibe~A. {de Jong}.
\newblock ``Hybrid {{Quantum-Classical Hierarchy}} for {{Mitigation}} of
  {{Decoherence}} and {{Determination}} of {{Excited States}}''.
\newblock \href{https://dx.doi.org/10.1103/PhysRevA.95.042308}{Physical Review
  A {\bf 95}, 042308}~(2017).
\newblock  \href{http://arxiv.org/abs/1603.05681}{arXiv:1603.05681}.

\bibitem{mccleanDecodingQuantumErrors2020}
Jarrod~R. McClean, Zhang Jiang, Nicholas~C. Rubin, Ryan Babbush, and Hartmut
  Neven.
\newblock ``Decoding quantum errors with subspace expansions''.
\newblock \href{https://dx.doi.org/10.1038/s41467-020-14341-w}{Nature
  Communications {\bf 11}, 636}~(2020).

\bibitem{yoshiokaGeneralizedQuantumSubspace2022a}
Nobuyuki Yoshioka, Hideaki Hakoshima, Yuichiro Matsuzaki, Yuuki Tokunaga,
  Yasunari Suzuki, and Suguru Endo.
\newblock ``Generalized {{Quantum Subspace Expansion}}''.
\newblock \href{https://dx.doi.org/10.1103/PhysRevLett.129.020502}{Physical
  Review Letters {\bf 129}, 020502}~(2022).

\bibitem{yangDualGSEResourceefficientGeneralized2024}
Bo~Yang, Nobuyuki Yoshioka, Hiroyuki Harada, Shigeo Hakkaku, Yuuki Tokunaga,
  Hideaki Hakoshima, Kaoru Yamamoto, and Suguru Endo.
\newblock ``Dual-{{GSE}}: {{Resource-efficient Generalized Quantum Subspace
  Expansion}}''~(2024).
\newblock  \href{http://arxiv.org/abs/2309.14171}{arXiv:2309.14171}.

\bibitem{getelinaQuantumSubspaceExpansion2024}
Jo{\~a}o~C. Getelina, Prachi Sharma, Thomas Iadecola, Peter~P. Orth, and
  Yong-Xin Yao.
\newblock ``Quantum subspace expansion in the presence of hardware noise''.
\newblock \href{https://dx.doi.org/10.1063/5.0217294}{APL Quantum {\bf 1},
  036127}~(2024).

\bibitem{hugginsVirtualDistillationQuantum2021}
William~J. Huggins, Sam McArdle, Thomas~E. O'Brien, Joonho Lee, Nicholas~C.
  Rubin, Sergio Boixo, K.~Birgitta Whaley, Ryan Babbush, and Jarrod~R. McClean.
\newblock ``Virtual {{Distillation}} for {{Quantum Error Mitigation}}''.
\newblock \href{https://dx.doi.org/10.1103/PhysRevX.11.041036}{Physical Review
  X {\bf 11}, 041036}~(2021).

\bibitem{koczorExponentialErrorSuppression2021}
B{\'a}lint Koczor.
\newblock ``Exponential {{Error Suppression}} for {{Near-Term Quantum
  Devices}}''.
\newblock \href{https://dx.doi.org/10.1103/PhysRevX.11.031057}{Physical Review
  X {\bf 11}, 031057}~(2021).

\bibitem{temmeErrorMitigationShortdepth2017}
Kristan Temme, Sergey Bravyi, and Jay~M. Gambetta.
\newblock ``Error mitigation for short-depth quantum circuits''.
\newblock \href{https://dx.doi.org/10.1103/PhysRevLett.119.180509}{Physical
  Review Letters {\bf 119}, 180509}~(2017).
\newblock  \href{http://arxiv.org/abs/1612.02058}{arXiv:1612.02058}.

\bibitem{saadRatesConvergenceLanczos1980}
Y.~Saad.
\newblock ``On the {{Rates}} of {{Convergence}} of the {{Lanczos}} and the
  {{Block-Lanczos Methods}}''.
\newblock \href{https://dx.doi.org/10.1137/0717059}{SIAM Journal on Numerical
  Analysis {\bf 17}, 687--706}~(1980).

\bibitem{parrishQuantumFilterDiagonalization2019}
Robert~M. Parrish and Peter~L. McMahon.
\newblock ``Quantum {{Filter Diagonalization}}: {{Quantum Eigendecomposition}}
  without {{Full Quantum Phase Estimation}}''~(2019).
\newblock  \href{http://arxiv.org/abs/1909.08925}{arXiv:1909.08925}.

\bibitem{klymkoRealTimeEvolutionUltracompact2022}
Katherine Klymko, Carlos {Mejuto-Zaera}, Stephen~J. Cotton, Filip Wudarski,
  Miroslav Urbanek, Diptarka Hait, Martin {Head-Gordon}, K.~Birgitta Whaley,
  Jonathan Moussa, Nathan Wiebe, Wibe~A. {de Jong}, and Norm~M. Tubman.
\newblock ``Real-{{Time Evolution}} for {{Ultracompact Hamiltonian
  Eigenstates}} on {{Quantum Hardware}}''.
\newblock \href{https://dx.doi.org/10.1103/PRXQuantum.3.020323}{PRX Quantum
  {\bf 3}, 020323}~(2022).

\bibitem{mottaDeterminingEigenstatesThermal2020}
Mario Motta, Chong Sun, Adrian T.~K. Tan, Matthew~J. O'Rourke, Erika Ye,
  Austin~J. Minnich, Fernando G. S.~L. Brand{\~a}o, and Garnet Kin-Lic Chan.
\newblock ``Determining eigenstates and thermal states on a quantum computer
  using quantum imaginary time evolution''.
\newblock \href{https://dx.doi.org/10.1038/s41567-019-0704-4}{Nature Physics
  {\bf 16}, 205--210}~(2020).

\bibitem{sekiQuantumPowerMethod2021}
Kazuhiro Seki and Seiji Yunoki.
\newblock ``Quantum {{Power Method}} by a {{Superposition}} of {{Time-Evolved
  States}}''.
\newblock \href{https://dx.doi.org/10.1103/PRXQuantum.2.010333}{PRX Quantum
  {\bf 2}, 010333}~(2021).

\bibitem{cortesQuantumKrylovSubspace2022}
Cristian~L. Cortes and Stephen~K. Gray.
\newblock ``Quantum {{Krylov}} subspace algorithms for ground and excited state
  energy estimation''.
\newblock \href{https://dx.doi.org/10.1103/PhysRevA.105.022417}{Physical Review
  A {\bf 105}, 022417}~(2022).
\newblock  \href{http://arxiv.org/abs/2109.06868}{arXiv:2109.06868}.

\bibitem{bhartiIterativeQuantumassistedEigensolver2021}
Kishor Bharti and Tobias Haug.
\newblock ``Iterative quantum-assisted eigensolver''.
\newblock \href{https://dx.doi.org/10.1103/PhysRevA.104.L050401}{Physical
  Review A {\bf 104}, L050401}~(2021).

\bibitem{leeSamplingErrorAnalysis2024}
Gwonhak Lee, Dongkeun Lee, and Joonsuk Huh.
\newblock ``Sampling {{Error Analysis}} in {{Quantum Krylov Subspace
  Diagonalization}}''.
\newblock \href{https://dx.doi.org/10.22331/q-2024-09-19-1477}{Quantum {\bf 8},
  1477}~(2024).
\newblock  \href{http://arxiv.org/abs/2307.16279}{arXiv:2307.16279}.

\bibitem{kirbyExactEfficientLanczos2023}
William Kirby, Mario Motta, and Antonio Mezzacapo.
\newblock ``Exact and efficient {{Lanczos}} method on a quantum computer''.
\newblock \href{https://dx.doi.org/10.22331/q-2023-05-23-1018}{Quantum {\bf 7},
  1018}~(2023).
\newblock  \href{http://arxiv.org/abs/2208.00567}{arXiv:2208.00567}.

\bibitem{low2019hamiltonian}
Guang~Hao Low and Isaac~L. Chuang.
\newblock ``Hamiltonian {S}imulation by {Q}ubitization''.
\newblock \href{https://dx.doi.org/10.22331/q-2019-07-12-163}{{Quantum} {\bf
  3}, 163}~(2019).

\bibitem{andersonSolvingLatticeGauge2025}
Lewis~W. Anderson, Martin Kiffner, Tom O'Leary, Jason Crain, and Dieter Jaksch.
\newblock ``Solving lattice gauge theories using the quantum {{Krylov}}
  algorithm and qubitization''.
\newblock \href{https://dx.doi.org/10.22331/q-2025-03-25-1669}{Quantum {\bf 9},
  1669}~(2025).
\newblock  \href{http://arxiv.org/abs/2403.08859}{arXiv:2403.08859}.

\bibitem{paige1971computation}
Christopher~Conway Paige.
\newblock ``The computation of eigenvalues and eigenvectors of very large
  sparse matrices.''.
\newblock PhD thesis.
\newblock University of London.
\newblock ~(1971).

\bibitem{epperlyTheoryQuantumSubspace2022}
Ethan~N. Epperly, Lin Lin, and Yuji Nakatsukasa.
\newblock ``A theory of quantum subspace diagonalization''.
\newblock \href{https://dx.doi.org/10.1137/21M145954X}{SIAM Journal on Matrix
  Analysis and Applications {\bf 43}, 1263--1290}~(2022).
\newblock  \href{http://arxiv.org/abs/2110.07492}{arXiv:2110.07492}.

\bibitem{kirbyAnalysisQuantumKrylov2024}
William Kirby.
\newblock ``Analysis of quantum {{Krylov}} algorithms with errors''.
\newblock \href{https://dx.doi.org/10.22331/q-2024-08-29-1457}{Quantum {\bf 8},
  1457}~(2024).
\newblock  \href{http://arxiv.org/abs/2401.01246}{arXiv:2401.01246}.

\bibitem{parlettSymmetricEigenvalueProblem1998}
Beresford~N Parlett.
\newblock ``The symmetric eigenvalue problem''.
\newblock \href{https://dx.doi.org/10.1137/1.9781611971163}{SIAM}. ~(1998).

\bibitem{kimEvidenceUtilityQuantum2023}
Youngseok Kim, Andrew Eddins, Sajant Anand, Ken~Xuan Wei, Ewout {van den Berg},
  Sami Rosenblatt, Hasan Nayfeh, Yantao Wu, Michael Zaletel, Kristan Temme, and
  Abhinav Kandala.
\newblock ``Evidence for the utility of quantum computing before fault
  tolerance''.
\newblock \href{https://dx.doi.org/10.1038/s41586-023-06096-3}{Nature {\bf
  618}, 500--505}~(2023).

\bibitem{nandkishoreManyBodyLocalizationThermalization2015}
Rahul Nandkishore and David~A. Huse.
\newblock ``Many-{{Body Localization}} and {{Thermalization}} in {{Quantum
  Statistical Mechanics}}''.
\newblock
  \href{https://dx.doi.org/10.1146/annurev-conmatphys-031214-014726}{Annual
  Review of Condensed Matter Physics {\bf 6}, 15--38}~(2015).

\bibitem{childsFirstQuantumSimulation2018}
Andrew~M. Childs, Dmitri Maslov, Yunseong Nam, Neil~J. Ross, and Yuan Su.
\newblock ``Toward the first quantum simulation with quantum speedup''.
\newblock \href{https://dx.doi.org/10.1073/pnas.1801723115}{Proceedings of the
  National Academy of Sciences {\bf 115}, 9456--9461}~(2018).
\newblock  \href{http://arxiv.org/abs/1711.10980}{arXiv:1711.10980}.

\bibitem{luitzManybodyLocalizationEdge2015}
David~J. Luitz, Nicolas Laflorencie, and Fabien Alet.
\newblock ``Many-body localization edge in the random-field {{Heisenberg}}
  chain''.
\newblock \href{https://dx.doi.org/10.1103/PhysRevB.91.081103}{Physical Review
  B {\bf 91}, 081103}~(2015).
\newblock  \href{http://arxiv.org/abs/1411.0660}{arXiv:1411.0660}.

\bibitem{szaboModernQuantumChemistry1996}
Attila Szabo and Neil~S Ostlund.
\newblock ``Modern quantum chemistry: Introduction to advanced electronic
  structure theory''.
\newblock Dover Publications. New York~(1996).

\bibitem{seeleyBravyiKitaevTransformationQuantum2012}
Jacob~T. Seeley, Martin~J. Richard, and Peter~J. Love.
\newblock ``The {{Bravyi-Kitaev}} transformation for quantum computation of
  electronic structure''.
\newblock \href{https://dx.doi.org/10.1063/1.4768229}{The Journal of Chemical
  Physics {\bf 137}, 224109}~(2012).
\newblock  \href{http://arxiv.org/abs/1208.5986}{arXiv:1208.5986}.

\bibitem{sun2018pyscf}
Qiming Sun, Timothy~C Berkelbach, Nick~S Blunt, George~H Booth, Sheng Guo,
  Zhendong Li, Junzi Liu, James~D McClain, Elvira~R Sayfutyarova, Sandeep
  Sharma, et~al.
\newblock ``{{PySCF}}: The {{Python-based}} simulations of chemistry
  framework''.
\newblock \href{https://dx.doi.org/10.1002/wcms.1340}{Wiley Interdisciplinary
  Reviews: Computational Molecular Science {\bf 8}, e1340}~(2018).

\bibitem{lehoucq1996deflation}
Richard~B Lehoucq and Danny~C Sorensen.
\newblock ``Deflation techniques for an implicitly restarted {{Arnoldi}}
  iteration''.
\newblock \href{https://dx.doi.org/10.1137/S0895479895281484}{SIAM Journal on
  Matrix Analysis and Applications {\bf 17}, 789--821}~(1996).

\bibitem{DAVIDSON197587}
Ernest~R. Davidson.
\newblock ``The iterative calculation of a few of the lowest eigenvalues and
  corresponding eigenvectors of large real-symmetric matrices''.
\newblock \href{https://dx.doi.org/10.1016/0021-9991(75)90065-0}{Journal of
  Computational Physics {\bf 17}, 87--94}~(1975).

\bibitem{tkachenkoQuantumDavidsonAlgorithm2024}
Nikolay~V Tkachenko, Lukasz Cincio, Alexander~I Boldyrev, Sergei Tretiak,
  Pavel~A Dub, and Yu~Zhang.
\newblock ``Quantum {{Davidson}} algorithm for excited states''.
\newblock \href{https://dx.doi.org/10.1088/2058-9565/ad3a97}{Quantum Science
  and Technology {\bf 9}, 035012}~(2024).

\bibitem{berthusenMultireferenceQuantumDavidson2024}
Noah Berthusen, Faisal Alam, and Yu~Zhang.
\newblock ``Multi-reference {{Quantum Davidson Algorithm}} for {{Quantum
  Dynamics}}''~(2024).
\newblock  \href{http://arxiv.org/abs/2406.08675}{arXiv:2406.08675}.

\bibitem{boydHighdimensionalSubspaceExpansion2025}
Gregory Boyd, B{\'a}lint Koczor, and Zhenyu Cai.
\newblock ``High-dimensional subspace expansion using classical shadows''.
\newblock \href{https://dx.doi.org/10.1103/PhysRevA.111.022423}{Physical Review
  A {\bf 111}, 022423}~(2025).

\bibitem{bluvsteinLogicalQuantumProcessor2023}
Dolev Bluvstein, Simon~J. Evered, Alexandra~A. Geim, Sophie~H. Li, Hengyun
  Zhou, Tom Manovitz, Sepehr Ebadi, Madelyn Cain, Marcin Kalinowski, Dominik
  Hangleiter, J.~Pablo~Bonilla Ataides, Nishad Maskara, Iris Cong, Xun Gao,
  Pedro~Sales Rodriguez, Thomas Karolyshyn, Giulia Semeghini, Michael~J.
  Gullans, Markus Greiner, Vladan Vuletic, and Mikhail~D. Lukin.
\newblock ``Logical quantum processor based on reconfigurable atom arrays''.
\newblock \href{https://dx.doi.org/10.1038/s41586-023-06927-3}{Nature}~(2023).
\newblock  \href{http://arxiv.org/abs/2312.03982}{arXiv:2312.03982}.

\bibitem{acharyaSuppressingQuantumErrors2023}
Rajeev Acharya, Igor Aleiner, Richard Allen, Trond~I. Andersen, Markus Ansmann,
  Frank Arute, Kunal Arya, Abraham Asfaw, Juan Atalaya, Ryan Babbush, Dave
  Bacon, Joseph~C. Bardin, Joao Basso, Andreas Bengtsson, Sergio Boixo, Gina
  Bortoli, Alexandre Bourassa, Jenna Bovaird, Leon Brill, Michael Broughton,
  Bob~B. Buckley, David~A. Buell, Tim Burger, Brian Burkett, Nicholas Bushnell,
  Yu~Chen, Zijun Chen, Ben Chiaro, Josh Cogan, Roberto Collins, Paul Conner,
  William Courtney, Alexander~L. Crook, Ben Curtin, Dripto~M. Debroy, Alexander
  Del Toro~Barba, Sean Demura, Andrew Dunsworth, Daniel Eppens, Catherine
  Erickson, Lara Faoro, Edward Farhi, Reza Fatemi, Leslie Flores~Burgos,
  Ebrahim Forati, Austin~G. Fowler, Brooks Foxen, William Giang, Craig Gidney,
  Dar Gilboa, Marissa Giustina, Alejandro Grajales~Dau, Jonathan~A. Gross,
  Steve Habegger, Michael~C. Hamilton, Matthew~P. Harrigan, Sean~D. Harrington,
  Oscar Higgott, Jeremy Hilton, Markus Hoffmann, Sabrina Hong, Trent Huang,
  Ashley Huff, William~J. Huggins, Lev~B. Ioffe, Sergei~V. Isakov, Justin
  Iveland, Evan Jeffrey, Zhang Jiang, Cody Jones, Pavol Juhas, Dvir Kafri,
  Kostyantyn Kechedzhi, Julian Kelly, Tanuj Khattar, Mostafa Khezri, M{\'a}ria
  Kieferov{\'a}, Seon Kim, Alexei Kitaev, Paul~V. Klimov, Andrey~R. Klots,
  Alexander~N. Korotkov, Fedor Kostritsa, John~Mark Kreikebaum, David Landhuis,
  Pavel Laptev, Kim-Ming Lau, Lily Laws, Joonho Lee, Kenny Lee, Brian~J.
  Lester, Alexander Lill, Wayne Liu, Aditya Locharla, Erik Lucero, Fionn~D.
  Malone, Jeffrey Marshall, Orion Martin, Jarrod~R. McClean, Trevor McCourt,
  Matt McEwen, Anthony Megrant, Bernardo Meurer~Costa, Xiao Mi, Kevin~C. Miao,
  Masoud Mohseni, Shirin Montazeri, Alexis Morvan, Emily Mount, Wojciech
  Mruczkiewicz, Ofer Naaman, Matthew Neeley, Charles Neill, Ani Nersisyan,
  Hartmut Neven, Michael Newman, Jiun~How Ng, Anthony Nguyen, Murray Nguyen,
  Murphy~Yuezhen Niu, Thomas~E. O'Brien, Alex Opremcak, John Platt, Andre
  Petukhov, Rebecca Potter, Leonid~P. Pryadko, Chris Quintana, Pedram Roushan,
  Nicholas~C. Rubin, Negar Saei, Daniel Sank, Kannan Sankaragomathi, Kevin~J.
  Satzinger, Henry~F. Schurkus, Christopher Schuster, Michael~J. Shearn, Aaron
  Shorter, Vladimir Shvarts, Jindra Skruzny, Vadim Smelyanskiy, W.~Clarke
  Smith, George Sterling, Doug Strain, Marco Szalay, Alfredo Torres, Guifre
  Vidal, Benjamin Villalonga, Catherine Vollgraff~Heidweiller, Theodore White,
  Cheng Xing, Z.~Jamie Yao, Ping Yeh, Juhwan Yoo, Grayson Young, Adam Zalcman,
  Yaxing Zhang, Ningfeng Zhu, and {Google Quantum AI}.
\newblock ``Suppressing quantum errors by scaling a surface code logical
  qubit''.
\newblock \href{https://dx.doi.org/10.1038/s41586-022-05434-1}{Nature {\bf
  614}, 676--681}~(2023).

\bibitem{krinnerRealizingRepeatedQuantum2022a}
Sebastian Krinner, Nathan Lacroix, Ants Remm, Agustin Di~Paolo, Elie Genois,
  Catherine Leroux, Christoph Hellings, Stefania Lazar, Francois Swiadek,
  Johannes Herrmann, Graham~J. Norris, Christian~Kraglund Andersen, Markus
  M{\"u}ller, Alexandre Blais, Christopher Eichler, and Andreas Wallraff.
\newblock ``Realizing {{Repeated Quantum Error Correction}} in a
  {{Distance-Three Surface Code}}''.
\newblock \href{https://dx.doi.org/10.1038/s41586-022-04566-8}{Nature {\bf
  605}, 669--674}~(2022).
\newblock  \href{http://arxiv.org/abs/2112.03708}{arXiv:2112.03708}.

\bibitem{ryan-andersonImplementingFaulttolerantEntangling2022a}
C.~{Ryan-Anderson}, N.~C. Brown, M.~S. Allman, B.~Arkin, G.~{Asa-Attuah},
  C.~Baldwin, J.~Berg, J.~G. Bohnet, S.~Braxton, N.~Burdick, J.~P. Campora,
  A.~Chernoguzov, J.~Esposito, B.~Evans, D.~Francois, J.~P. Gaebler, T.~M.
  Gatterman, J.~Gerber, K.~Gilmore, D.~Gresh, A.~Hall, A.~Hankin, J.~Hostetter,
  D.~Lucchetti, K.~Mayer, J.~Myers, B.~Neyenhuis, J.~Santiago, J.~Sedlacek,
  T.~Skripka, A.~Slattery, R.~P. Stutz, J.~Tait, R.~Tobey, G.~Vittorini,
  J.~Walker, and D.~Hayes.
\newblock ``Implementing {{Fault-tolerant Entangling Gates}} on the
  {{Five-qubit Code}} and the {{Color Code}}''~(2022).
\newblock  \href{http://arxiv.org/abs/2208.01863}{arXiv:2208.01863}.

\bibitem{richardsUniversityOxfordAdvanced2015}
Andrew Richards.
\newblock ``University of {{Oxford Advanced Research Computing}}''~(2015).

\bibitem{hugginsNonOrthogonalVariationalQuantum2020}
William~J. Huggins, Joonho Lee, Unpil Baek, Bryan O'Gorman, and K.~Birgitta
  Whaley.
\newblock ``A {{Non-Orthogonal Variational Quantum Eigensolver}}''.
\newblock \href{https://dx.doi.org/10.1088/1367-2630/ab867b}{New Journal of
  Physics {\bf 22}, 073009}~(2020).
\newblock  \href{http://arxiv.org/abs/1909.09114}{arXiv:1909.09114}.

\end{thebibliography}

\onecolumn
\appendix

\section{Additional numerical results} \label{app:add_num}
\begin{figure}
    \centering
    \includegraphics[width=\textwidth]{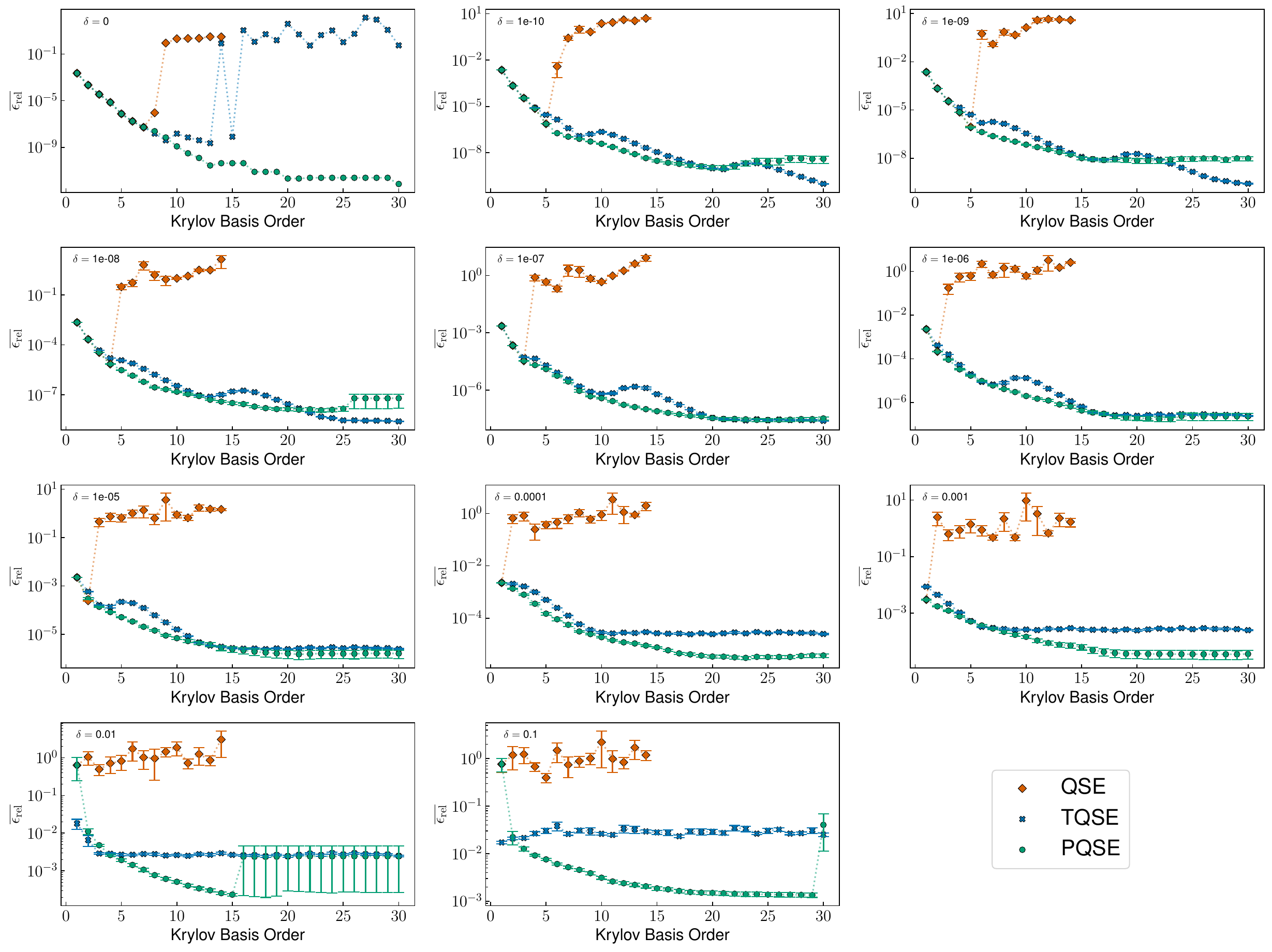}
    \caption{Average relative error in estimate of ground state energy for QSE with a Krylov basis generated using Hamiltonian powers: unmodified (QSE), thresholded (TQSE), partitioned (PQSE) with increasing $\delta$. The target system is the 10 qubit periodic disordered Heisenberg model with $J=0.1$ and $h=1$ for increasing Gaussian noise strength.}
    \label{fig:all_noise_err_decay_sr}
\end{figure}

\begin{figure}
    \centering
    \includegraphics[width=\textwidth]{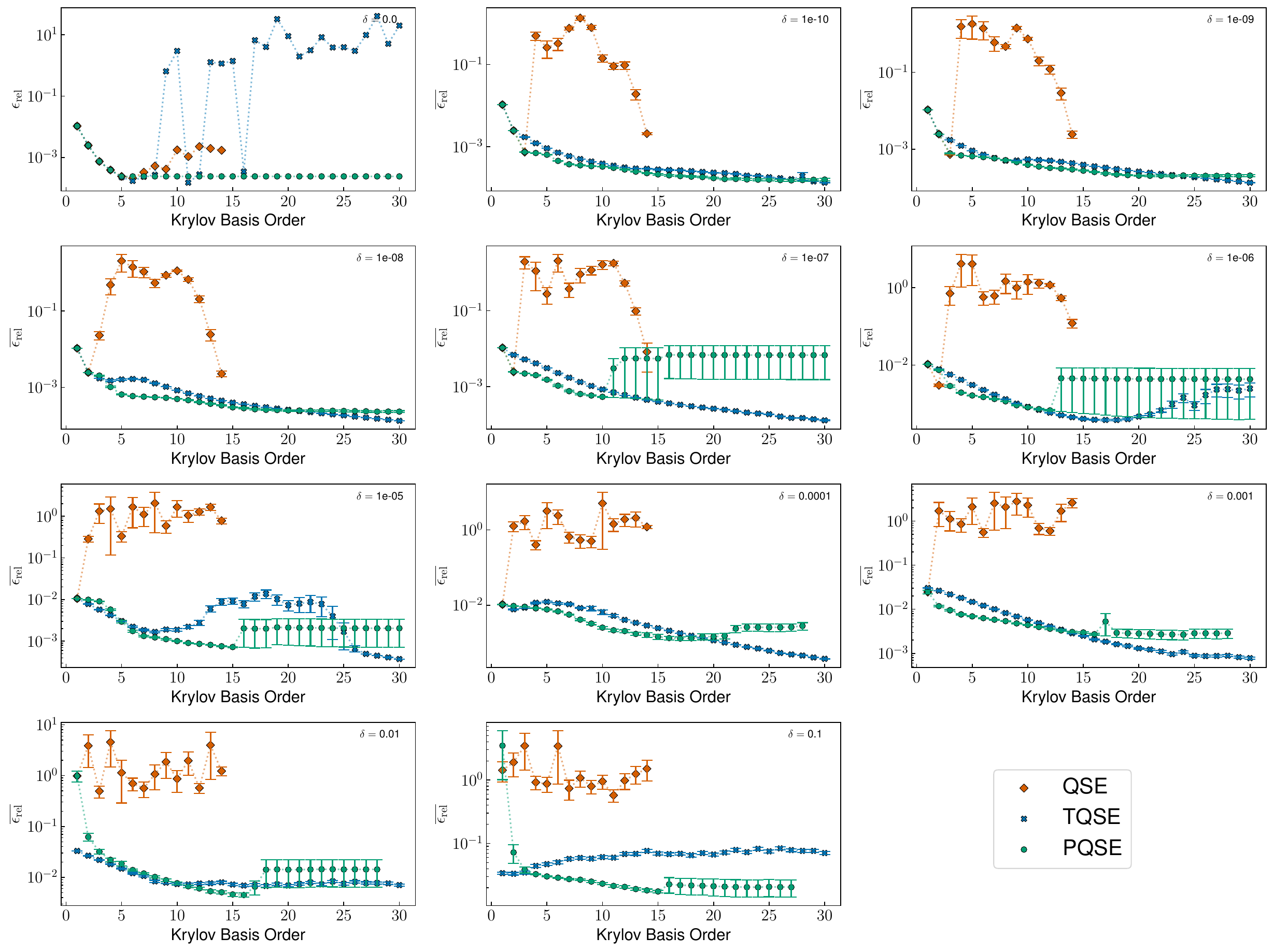}
    \caption{Average relative error in estimate of ground state energy for QSE with a Krylov basis generated using Hamiltonian powers: unmodified (QSE), thresholded (TQSE), partitioned (PQSE) with increasing $\delta$. The target system is H$_6$ in a STO-3G basis with a parity mapping applied in equilibrium configuration.}
    \label{fig:all_noise_err_decay_h6}
\end{figure}

Here we provide additional data illustrating the average relative error with increasing maximum allowed Krylov basis order associated with different approaches to QSE. We present a wider range of noise strengths $\delta$ than considered in the main text. Additional results for the spin ring example are in Fig. \ref{fig:all_noise_err_decay_sr} and Hydrogen chain example in Fig. \ref{fig:all_noise_err_decay_h6}. A detailed description of the setup for these examples can be found in Sec.~(\ref{sec:num}) in the main text. We highlight that the results shown in Fig. \ref{fig:min_ave_error} in the main text can be recognised by taking the minimum average error across all Krylov basis orders for each noise strength $\delta > 0$. Fig. \ref{fig:min_ave_error}a connects to Fig. \ref{fig:all_noise_err_decay_sr} and Fig. \ref{fig:min_ave_error}b to Fig. \ref{fig:all_noise_err_decay_h6}.

\begin{figure}
    \centering
    \includegraphics[width=0.5\linewidth]{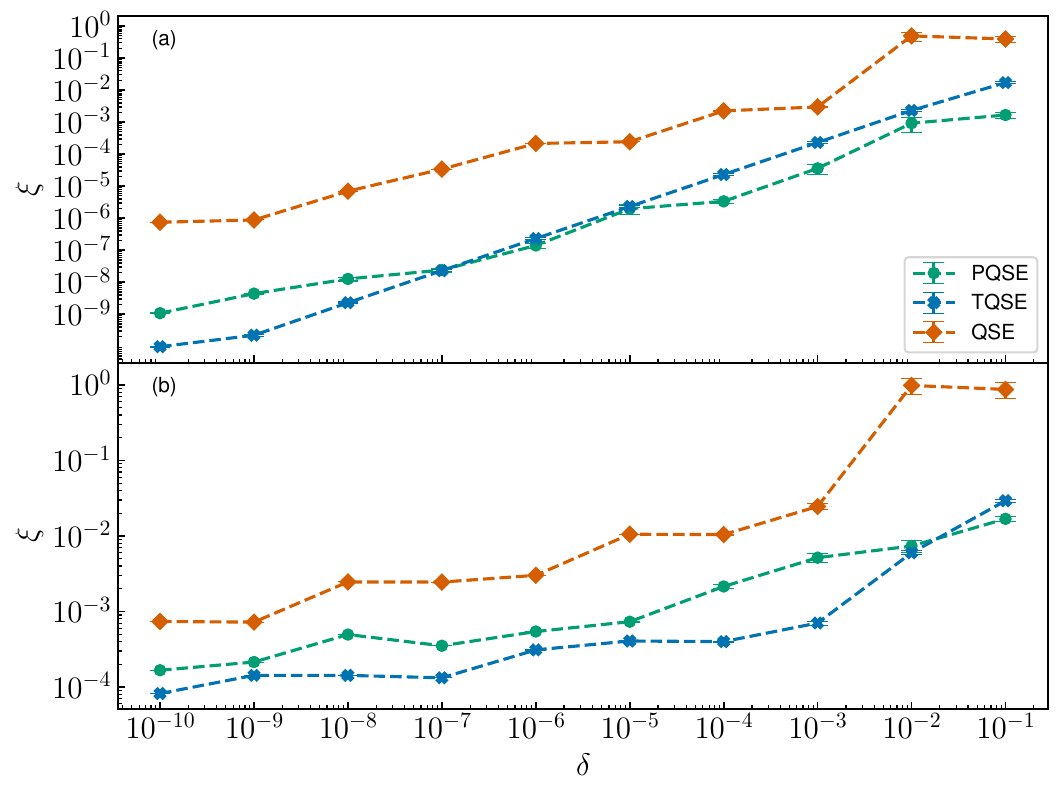}
    \caption{Minimum average relative error $\xi$ in the estimate of ground state energy  for QSE with a Krylov basis generated using Hamiltonian powers: unmodified (QSE), thresholded (TQSE), partitioned (PQSE). Here PQSE uses a more economical basis selection criterion, comparing solution energy instead of energy variance at a terminating Krylov basis order.
    (a) $\xi$ as a function of noise strength $\delta$ for a  10 qubit periodic disordered Heisenberg model with $J/h=0.1$.  (b)  $\xi$ as a function of noise strength $\delta$ for H$_6$ in a STO-3G basis with a parity mapping applied in equilibrium configuration. Error bars show the standard error.}
    \label{fig:min_ave_error_alt}
\end{figure}

Additionally, we present the minimum average error $\xi$ with increasing noise strength when PQSE is using an alternative basis selection criterion.
Here, once PQSE has constructed a state involving a terminating Hamiltonian power: $H^{\kroF - 1}$, it will use the state's energy squared, computed using the procedure in Appendix~(\ref{app:recon}), instead of the energy variance as a metric to compare with other possible solutions.
We present the corresponding results in Figure~\ref{fig:min_ave_error_alt}, finding that the performance of PQSE here closely resembles that in the main text Figure~(\ref{fig:min_ave_error}).

\newpage

\section{Generating a Krylov Basis through Real Time Evolution} \label{app:rte}
In this section we present the results for our spin ring and molecular examples using a Krylov basis generated in by real time evolution.
This approach is more suited to near term quantum computers as it typically involves shallower quantum circuits than those needed to generate a Krylov basis with Hamiltonian powers. 
The subspace is generated by real time evolution (RTE)
\begin{align} \label{eq:rte_basis}
\tilde{\mathcal{K}}_{\kroF}(H,\ket{\phi_0}) = \text{span} \{ e^{- \mathrm{i} H t_j} \ket{\phi_0} \}_{j=0}^{R-1},
\end{align}
for some choice of evolution times $\{t_j\}$, 
where $t_j = j \text{d}t $ for a choice of timestep $\text{d}t$.
We consider $\text{d}t = \pi / \lVert H \rVert$, where $\lVert \cdot \rVert$ is the spectral norm.
This provides optimal convergence to the classical Krylov subspace \cite{epperlyTheoryQuantumSubspace2022}.
The QSE matrices $\mathbf{H}$ and $\mathbf{S}$ then have elements
\begin{subequations}
\label{eq:rte_mat_ele}
\begin{align} 
    H_{ij} &= \braket{\psi_i \vert H \vert \psi_j}\,, \\
S_{ij} &= \braket{\psi_i \vert \psi_j}\,, \\
i,j &\in \{0,1,\ldots,\kroF-1\},
\end{align}
\end{subequations}
where $\ket{\psi_j} = e^{- \mathrm{i} j \text{d}t H} \ket{\phi_0}$.

We consider the same numerical setup as described in the main text.
For TQSE we follow the same parameter scan procedure.
We highlight two differences to the main text numerical experiments necessitated by the change of basis
\begin{enumerate}
    \item \textit{Finite shot noise implementation.}
    With the RTE basis we have to estimate the finite shot error involved in computing matrix elements of the form $\bra{\psi_i} H \ket{\psi_j}$ and $\braket{\psi_i \lvert \psi_j}$.
    We add noise in the same way as the power basis experiments by assuming the measurement error in estimating $\bra{\psi_i} H \ket{\psi_j}$ is well approximated by the error in estimating $\bra{\psi_j} H \ket{\psi_j}$.
    This is motivated by existing procedures for estimating off-diagonal matrix elements e.g. see Section 2.1 of \cite{hugginsNonOrthogonalVariationalQuantum2020}.
    For $S_{ij} = \braket{\psi_i \lvert \psi_j}$ we draw noise from a distribution of width $\delta$ instead of $\delta \sqrt{\text{Var}[S_{ij}]}$.

    \item \textit{Variance criterion matrix elements}.
    For consistency with the rest of the paper we use the variance minimization criterion for PQSE again.
    However, we note that we no longer get the matrix elements required to compute solution energy variance ``for free'', as $\bra{\psi_i} H^2 \ket{\psi_j}$ is not used in QSE. 
    We add noise in the same way to the other matrix elements.  
\end{enumerate}

\begin{figure}
    \centering
    \begin{tabular}{c}
     (a)\includegraphics[width=0.5\linewidth]{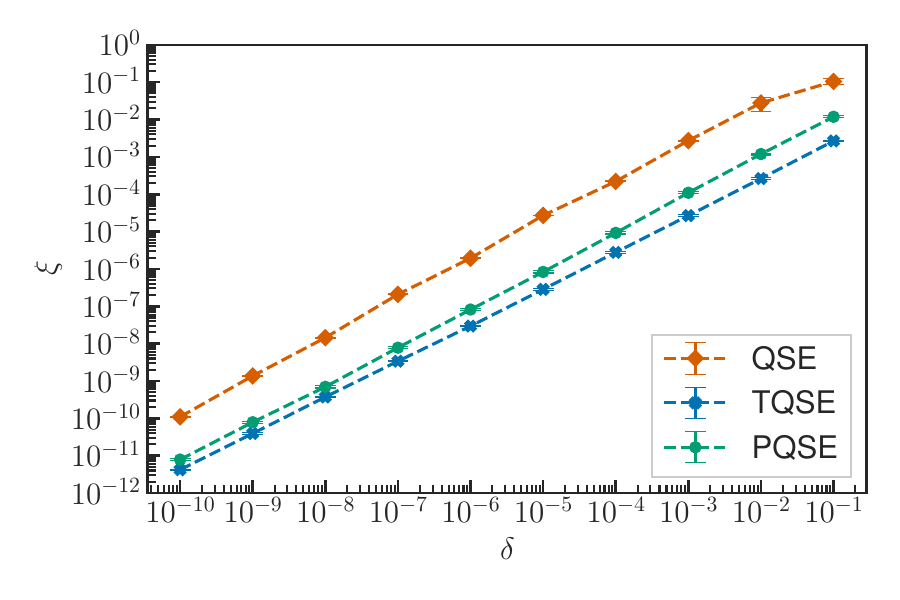}  \\
     (b)\includegraphics[width=0.5\linewidth]{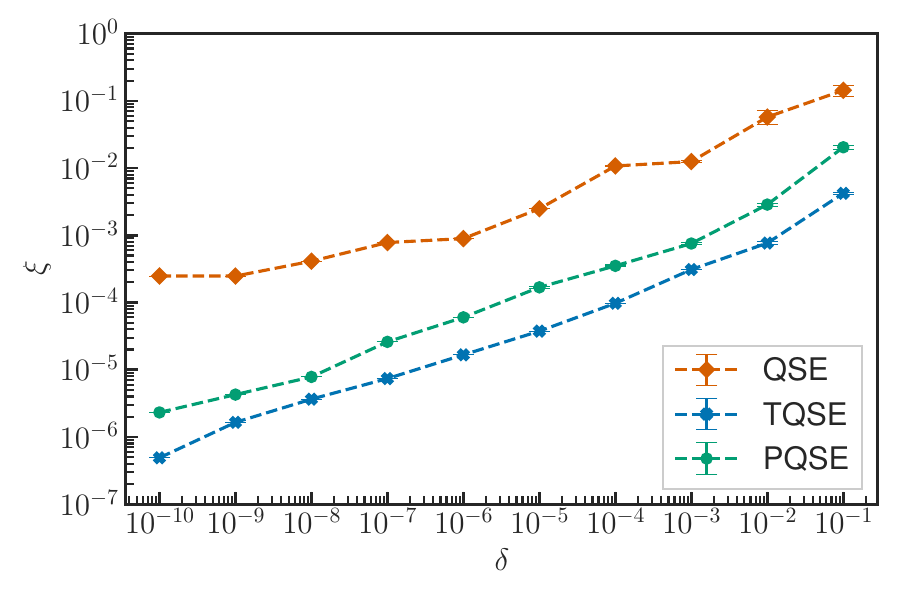}
    \end{tabular}
    \caption{Minimum average relative error $\xi$ in the estimate of ground state energy for QSE with a Krylov basis generated using real time evolution: unmodified (QSE), thresholded (TQSE), partitioned (PQSE). 
    (a) $\xi$ as a function of noise strength $\delta$ for a  10 qubit periodic disordered Heisenberg model with $J/h=0.1$.  (b)  $\xi$ as a function of noise strength $\delta$ for H$_6$ in a STO-3G basis with a parity mapping applied in equilibrium configuration. Error bars show the standard error.}
    \label{fig:min_ave_error_rte}
\end{figure}

\begin{figure}
    \centering
    \includegraphics[width=\linewidth]{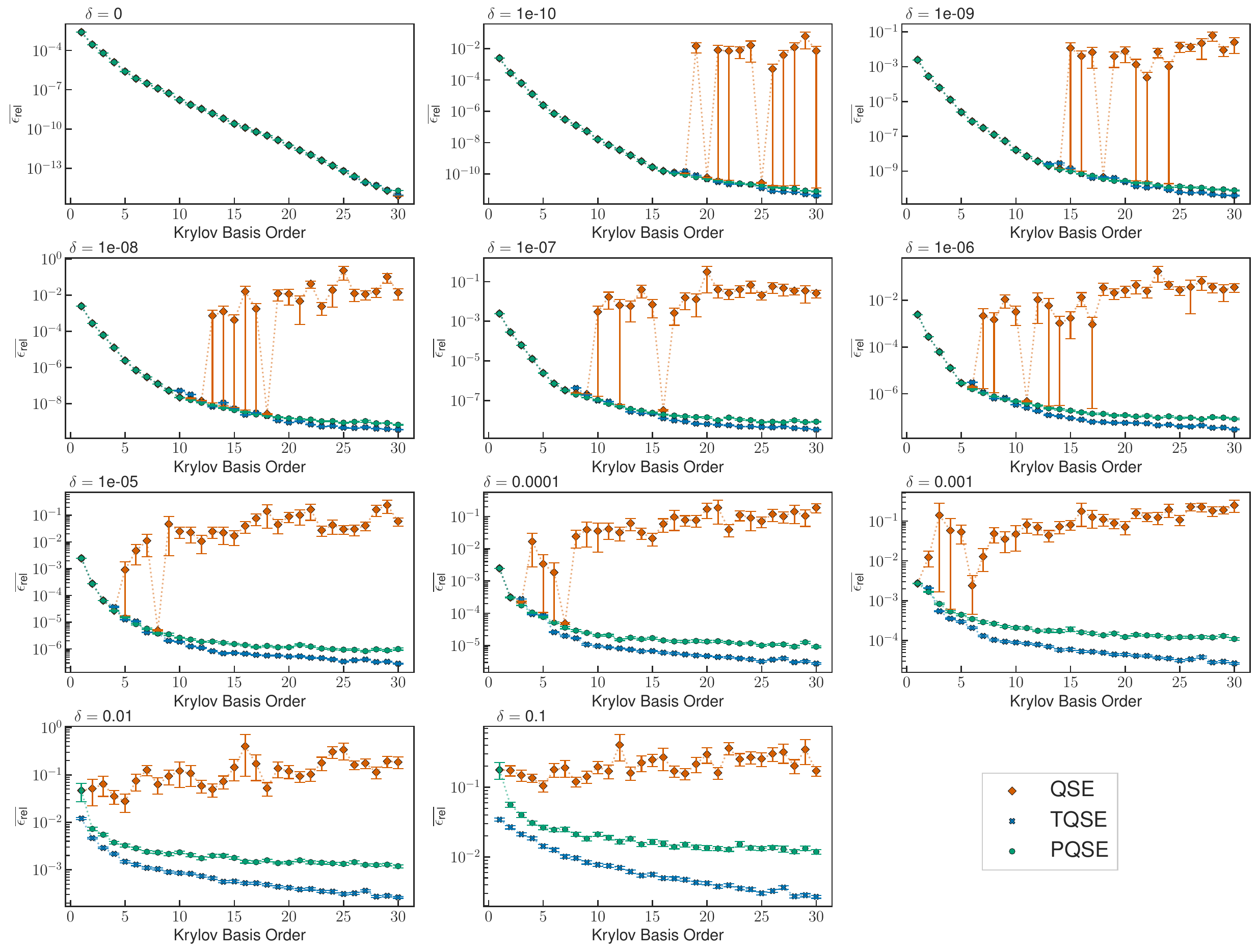}
    \caption{Average relative error in estimate of ground state energy for QSE with a Krylov basis generated using real time evolution: unmodified (QSE), thresholded (TQSE), partitioned (PQSE) with increasing $\delta$. The target system is the 10 qubit periodic disordered Heisenberg model with $J=0.1$ and $h=1$ for increasing Gaussian noise strength.}
    \label{fig:all_noise_err_decay_sr_rte}
\end{figure}

\begin{figure}
    \centering
    \includegraphics[width=\linewidth]{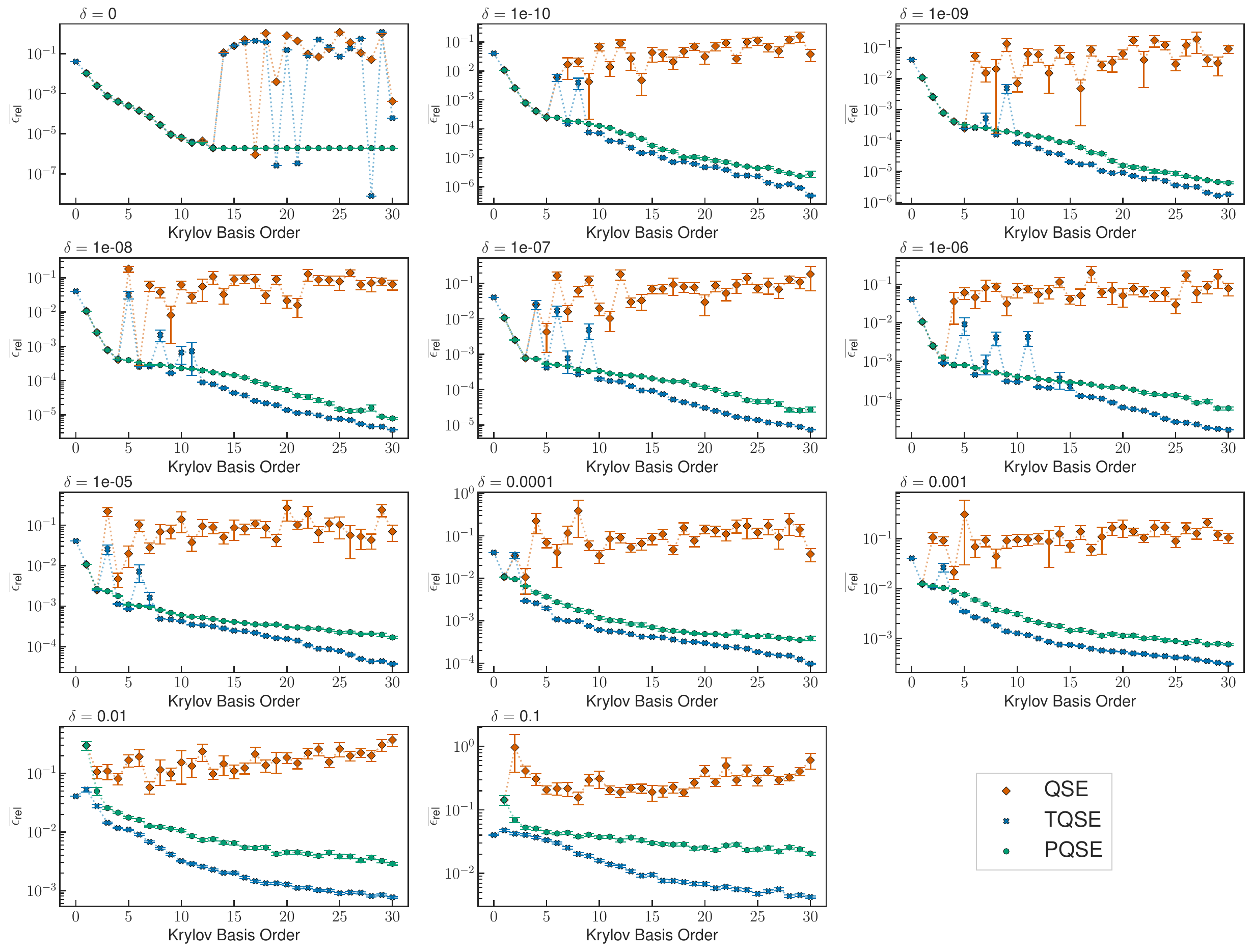}
    \caption{Average relative error in estimate of ground state energy for QSE with a Krylov basis using real time evolution: unmodified (QSE), thresholded (TQSE), partitioned (PQSE) with increasing $\delta$. The target system is H$_6$ in a STO-3G basis with a parity mapping applied in equilibrium configuration.}
    \label{fig:all_noise_err_decay_h6_rte}
\end{figure}

We now present the results using a RTE basis.
In Fig. \ref{fig:min_ave_error_rte} we include the minimum average error at each noise strength for both numerical examples and each QSE approach.
In Fig. \ref{fig:min_ave_error_rte}(a), for the spin ring example, we see that PQSE and TQSE have very similar accuracy. TQSE has a slight advantage at $\delta = 10^{-10}$, which increases up to around half an order of magnitude improvement at $\delta = 10^{-1}$. 
In Fig. \ref{fig:min_ave_error_rte}(b), for the H$_6$ example, we see that PQSE and TQSE have similar accuracy. TQSE has an advantage of approximately half an order of magnitude for all noise strengths. 
The full set of results used to produce these plots are included in Figs. \ref{fig:all_noise_err_decay_sr_rte} and \ref{fig:all_noise_err_decay_h6_rte} for the spin ring and molecular examples respectively.
These results highlight that PQSE is also competitive with parameter-optimized TQSE for the RTE-generated Krylov basis. This is notable since the TQSE parameter optimization used knowledge of the exact ground state energy. 
Although these results worked with exact time evolution and in practice some approximation error will be present, we consider this as positive motivation for applying PQSE to the output of near-term experiments.

\newpage

\section{Updating the reference state} \label{app:class}

\subsection{Hamiltonian power Krylov basis} \label{app:recon_ham}

Our algorithm uses the lowest energy state of one PQSE iteration as a reference state for the next. Here we show how the need to physically prepare a new reference state at each iteration can be circumvented through classical post-processing. 
We will show 
how PQSE can be implemented with only the ability to prepare the reference state $\ket{\phi_0}$ which is used to construct the order $R$ Krylov subspace 
\begin{align}
 \ket{\phi_k} = H^k\ket{\phi_0},\quad   k\in\{0,\ldots ,\kroF-1\}\,.
 \label{eq:FullKrylov}
\end{align}
The key observation is that the matrix elements required to set up the generalised eigenvalue problem at the $P$th iteration can be expanded as a sum of the form 
\begin{align} \label{eq:inner_prod_decomp}
    H^{\{P\}}_{ij} & = \sum_{m = 1}^{N_H} o_m \bra{\phi_0}H^{i + j + m}\ket{\phi_0}\,,\\
    S^{\{P\}}_{ij} & = \sum_{m = 0}^{N_S} o_m \bra{\phi_0}H^{i + j + m}\ket{\phi_0}\,,
\end{align}
where $o_m$ are updated at each iteration and $1 \leq N_H \leq 2R - 1$ and $0 \leq N_S \leq 2R - 2$. To illustrate this we consider obtaining the solution $\ket{\psi_1}$ from the first iteration, and the problem of constructing a new overlap matrix $S^{\{2\}}_{ij}$ for the next (with a similar derivation required to construct $H^{\{2\}}_{ij}$)
\begin{align*}
S^{\{2\}}_{ij} & = \bra{\psi_1} H^{i+j} \ket{\psi_1}\,,\\
& = \sum_{k,l = 0}^{r_1 - 1} c^{\text{opt}, *}_{r_1, k} c^{\text{opt}}_{r_1, l} \bra{\phi_0} H^{i+j+k+l} \ket{\phi_0}\,,
\end{align*}
where $c^{\text{opt}}_{r_1, l}$ are elements of the vector of optimal QSE coefficients $\textbf{c}^{\text{opt}}_{r_1}[\ket{\phi_0}]$. For a given $0 \leq m \leq 2 r_1 - 2$ values of $k$, $l$ satisfying $k+l=m$ will contribute multiple terms with the same expectation $\bra{\phi_0} H^{i+j+m} \ket{\phi_0}$. We can group these terms
\begin{align}
S^{\{2\}}_{ij} & = \sum_{m = 0}^{2r_1 - 2} \left( \sum_{\substack{k, l = 0: \\ k + l = m}}^{r_1 - 1} c^{\text{opt} *}_{r_1, k} c^{\text{opt}}_{r_1, l}\right) \bra{\phi_0} H^{i+j+m} \ket{\phi_0}\,,\\
& = \sum_{m = 0}^{2r_1 - 2} o_{m, r_1} \bra{\phi_0} H^{i+j+m} \ket{\phi_0}\,,
\end{align}
where 
\begin{align}
    o_{m,  r_1} =  \left( \sum_{\substack{k, l = 0: \\ k + l = m}}^{r_1 - 1} c^{\text{opt} *}_{r_1, k} c^{\text{opt}}_{r_1, l}\right).
\end{align} Having shown how the inner product coefficients appear in Eq.~(\ref{eq:inner_prod_decomp}), we now show how to update these coefficients after another iteration. If the QSE problem defined by $S^{\{2\}}_{ij}$ gives a solution $\ket{\psi_2} = \sum_{k = 0}^{r_2 - 1} c^{\text{opt}}_k H^k \ket{\psi_1}$, then at the next iteration
\begin{align}
    S^{\{3\}}_{ij} & = \bra{\psi_2} H^{i+j} \ket{\psi_2}\,,\\
     & = \sum_{k,l = 0}^{r_2 - 1} c^{\text{opt} *}_{r_2, k} c^{\text{opt}}_{r_2, l} \bra{\psi_1} H^{i+j+k+l} \ket{\psi_1}\,,\\
     & = \sum_{y = 0}^{2r_2 - 2} \left( \sum_{\substack{k, l = 0: \\ k + l = y}}^{r_2 - 1} c^{\text{opt} *}_{r_2, k} c^{\text{opt}}_{r_2, l} \right) \sum_{m = 0}^{2 r_1 - 2} o_{m, r_1} \bra{\phi_0} H^{i+j+y + m} \ket{\phi_0}\,,\\
     & = \sum_{y = 0}^{2r_2 - 2} \sum_{m = 0}^{2 r_1 - 2}  o_{y,  r_2} o_{m,  r_1} \bra{\phi_0} H^{i+j+y + m} \ket{\phi_0}\,,
\end{align}
where 
\begin{align}
    o_{y,  r_2} =  \left( \sum_{\substack{k, l = 0: \\ k + l = y}}^{r_2- 1} c^{\text{opt} *}_{r_2, k} c^{\text{opt}}_{r_2, l} \right).
\end{align}
Again, for a given $0 \leq m' \leq 2 r_1 + 2 r_2 - 4$ values of $y, m$ satisfying $y + m = m'$ will contribute multiple terms with the same expectation. Grouping these terms gives
\begin{align}
    S^{\{3\}}_{ij} & = \sum_{y = 0}^{2r_2 - 2} \sum_{m = 0}^{2 r_1 - 2}  o_{y, r_2} o_{m, r_1} \bra{\phi_0} H^{i+j+y + m} \ket{\phi_0}\,,\\
    & =  \sum_{m' = 0}^{2r_2 + 2 r_1 - 4} \left( \mathop{\sum_{y = 0}^{2r_2 - 2} \sum_{m = 0}^{2 r_1 - 2}}_{y + m = m'}  o_{y, r_2} o_{m, r_1} \right) \bra{\phi_0} H^{i+j+m'} \ket{\phi_0}\,,\\
    & = \sum_{m' = 0}^{2r_2 + 2 r_1 - 4} o_{m', \{r_1, r_2\}} \bra{\phi_0} H^{i+j+m'} \ket{\phi_0}\,,
\end{align}
where 
\begin{align} \label{eq:coeff_update}
o_{m', \{r_1, r_2\}} &= \mathop{\sum_{y = 0}^{2r_2 - 2} \sum_{m = 0}^{2 r_1 - 2}}_{y + m = m'}  o_{y, r_2} o_{m, r_1}.
\end{align}
This applies to constructing $\mathbf{S}^{\{P\}}$ at an arbitrary PQSE step as Eq. (\ref{eq:coeff_update}) can be used at each iteration after the first to update inner product coefficients $\{o_m\}$ using the optimal coefficients output from the most recent QSE problem $\textbf{c}^{\text{opt}}_{r_P}[\ket{\psi_{P-1}}]$. With each iteration, more inner products are used to construct new QSE problems. At the $P$th iteration there will be $N_P$ non-zero inner product coefficients where 
\begin{align} \label{eq:np}
    N_P = 2\sum_{i=1}^P r_i - 2P + 1\,.
\end{align}
In general, at the $P+1$th iteration the inner-product coefficients are updated as $\{ o_{m, \{\mathcal{R}\}} \}_{m=0}^{N_{P-1}} \rightarrow \{ o_{m' \{\mathcal{R}'\}}\}_{m'=0}^{N_{P}}$ where $\mathcal{R} = \{r_1, r_2, \dots , r_{P-1}\}$ and $\mathcal{R'} = \{r_1, r_2, \dots , r_{P}\}$ specify a sequence of Krylov basis orders. If the optimal QSE coefficients obtained at the $P$th iteration were $\textbf{c}^{\text{opt}}_{r_P}[\ket{\psi_{P-1}}]$, then the inner-product coefficients are updated using
\begin{align} \label{eq:inner_prod_coeff_update_general}
o_{m', \mathcal{R}'} &= \mathop{\sum_{y = 0}^{2r_P - 2} \sum_{m = 0}^{N_{P-1}}}_{y + m = m'}  o_{y, r_P} o_{m,  \mathcal{R}}\,,
\end{align}
where 
\begin{align} \label{eq:new_inner_prod_coeff_constr}
   o_{y, r_P}   =  \left( \sum_{\substack{k, l = 0: \\ k + l = y}}^{r_P - 1} c^{\text{opt} *}_{r_P, k} c^{\text{opt}}_{r_P, l} \right).
\end{align}
As PQSE has the constraint that 
\begin{align} \label{eq:cons}
  \sum_{i=1}^P r_i - P + 1 \leq R \,,
\end{align}
there will be at most $N_P = 2R - 1$ inner product coefficients. If we consider $S_{ij}  = \bra{\phi_0} H^{i+j} \ket{\phi_0} $ for single-step QSE with $0 \leq i,j \leq R - 1$, we see $N_P$ matches the number of unique inner products required to construct $\mathbf{S}$. 

\subsection{Resource cost} \label{app:cost}
Here we estimate the classical cost of PQSE. Our starting point is the observation that any sub-QSE problem with reference state $\ket{\psi}$ at the $P > 1$th PQSE iteration requires constructing matrices $\mathbf{H}^{\{P\}}$ and $\mathbf{S}^{\{P\}}$. The latter can be written
\begin{align} \label{eq:overlap_decomp_general}
    S^{\{P\}}_{ij} & = \bra{\psi}H^{i + j}\ket{\psi}\,,\\
    & = \sum_{m = 0}^{N_{P-1}} o_m \bra{\phi_0}H^{i + j + m}\ket{\phi_0}\,,  \label{eq:new_mat_ele}
\end{align}
where $0 \leq N_{P} \leq 2R - 2$ and $N_P$ is defined in Eq.~(\ref{eq:np}).
The cost of constructing  $S^{\{P+1\}}_{ij}$ at a new iteration comes from the cost of updating the inner-product coefficients $\{ o_{m, \mathcal{R}} \}_{m=0}^{N_{P-1}} \rightarrow \{ o_{m', \mathcal{R}'} \}_{m'=0}^{N_{P}}$. If the subspace dimension chosen at the $P$th iteration is $r_{P}$, then the coefficients are updated using Eqs.~(\ref{eq:inner_prod_coeff_update_general})~and~(\ref{eq:new_inner_prod_coeff_constr}).
Computing $\{ o_{y, r_P} \}_{y=0}^{2r_P - 2}$ in Eq.~(\ref{eq:new_inner_prod_coeff_constr}) can be done in $\mathcal{O}(r_P^2)$ basic arithmetic operations. Computing $\{ o_{m', \mathcal{R}'} \}_{m'=0}^{N_{P}}$ in Eq.~(\ref{eq:inner_prod_coeff_update_general}) requires $\mathcal{O}(r_P N_{P-1})$ basic arithmetic operations. The cost of updating the inner-product coefficients is therefore at most  $\mathcal{O}(R^2)$.

With the updated coefficients in hand, constructing an element of $\mathbf{S}^{\{P\}}$ using Eq. (\ref{eq:new_mat_ele}) requires $N_{P}$ basic steps. The same inner-product 
coefficients can be used to construct a single element of $\mathbf{H}^{\{P\}}$, which also requires $N_{P}$ basic steps. By making use of the Hankel structure of the 
$\mathbf{S}$ and $\mathbf{H}$ matrices, setting up the generalised eigenvalue problem in $\mathcal{K}_{r_{P+1}}(H, \ket{\psi})$ requires $4 r_{P+1} N_{P}$ operations. Constructing 
all possible subspaces for a given $r_{P+1}$ and solving the generalised eigenvalue problem with cost  $w r^3$ \cite{parlettSymmetricEigenvalueProblem1998}, 
where $w$ is a constant factor, has thus the overall cost 
\begin{align}
C(P) \leq R^2 + \sum\limits_{r_{P+1}=2}^{R-A}(4 r_{P+1} N_{P} + w r^3)\,,
\end{align}
Where $A = \sum_{i=1}^P r_i - P + 1$ limits $r_{P+1}$ in a way that satisfies the constraint in Eq. (\ref{eq:cons}).
Evaluating the sum shows that the dominant term scales like $R^4$. 
A full run of PQSE returns the sequence $\mathcal{R} = \{r_1,\ldots,r_B\}$ 
and thus requires $B$ times the steps detailed above. It follows that  the numerical cost for 
the classical part of PQSE is upper-bounded by $B R^4$.

\subsection{Calculating expectation values for a general Krylov basis} \label{app:recon}
PQSE returns $\ket{\psi_P}$ as an approximation to the ground state of $H$ in terms of expansion coefficients $\textbf{c}^{\text{opt}}_{r_{P-1}}[\ket{\phi_{P-1}}]$ relative to the Krylov basis with 
starting vector $\ket{\psi_{P-1}}$. 
We obtain the energy of this state $E_P$ when solving Eq.~(\ref{eq:gev}) in $\mathcal{K}_{r_P}(A,\ket{\psi_{P-1}})$ [see Sec.~\ref{background} and \ref{PQSE} in the main text]. 
In this section we show how to compute expectation values of observables in the state $\ket{\psi_P}$ in terms of expectation values  in the initial state $\ket{\phi_0}$ and a Krylov basis generating operator $A$. 
For example, this allows us to calculate the energy variance in $\ket{\psi_P}$.

The expected value of observable $\hat{O}$ after one PQSE iteration can be written

\begin{align} 
    \bra{\psi_1} \hat{O} 
    \ket{\psi_1} & = \sum_{k,l=0}^{r_1-1} c^{\text{opt}, *}_{r_1, k} c^{\text{opt}}_{r_1, l} \bra{\phi_0} (A^{k})^{\dagger} \hat{O} A^{l} \ket{\phi_0}\,, 
\end{align}

and after another PQSE iteration

\begin{align} 
    \bra{\psi_2} \hat{O} \ket{\psi_2} &= \sum_{k_2,l_2=0}^{r_2-1} c^{\text{opt}, *}_{r_2, k_2} c^{\text{opt}}_{r_2, l_2} \bra{\psi_1} (A^{k_2})^{\dagger} \hat{O} A^{l_2} \ket{\psi_1}\,, \\
    &= \sum_{k_2,l_2=0}^{r_2-1} \sum_{k_1,l_1=0}^{r_1-1} c^{\text{opt}, *}_{r_2, k_2} c^{\text{opt}, *}_{r_1, k_1}c^{\text{opt}}_{r_2, l_2}  c^{\text{opt}}_{r_1, l_1} \bra{\phi_0} (A^{k_1 + k_2})^{\dagger} \hat{O} A^{l_1 + l_2} \ket{\phi_0}\,.
\end{align}

For a given $0 \leq m, m' \leq r_1 + r_2 - 2$ values of $k_1$, $k_2$ satisfying $k_1+k_2=m$ and values of $l_1$, $l_2$ satisfying $l_1+l_2=m'$  will contribute multiple terms with the same expectation $\bra{\phi_0} (A^{m})^{\dagger} \hat{O} A^{m'}\ket{\phi_0}$. 
We can group these terms

\begin{align} \label{eq:obs_inner_prod_group}
    \bra{\psi_2} \hat{O} \ket{\psi_2} &= \sum_{m, m' = 0}^{r_1 + r_2 - 2} \left( \mathop{\sum_{k_2=0}^{r_2-1} \sum_{k_1,=0}^{r_1-1}}_{k_1 + k_2 = m} c^{\text{opt}, *}_{r_2, k_2} c^{\text{opt}, *}_{r_1, k_1} \right) \left( \mathop{\sum_{l_2=0}^{r_2-1} \sum_{l_1=0}^{r_1-1}}_{l_1 + l_2 = m'}  c^{\text{opt}}_{r_2, l_2}  c^{\text{opt}}_{r_1, l_1} \right) \bra{\phi_0} (A^{m})^{\dagger} \hat{O} A^{m'} \ket{\phi_0}\,.\\
    &= \sum_{m, m' = 0}^{r_1 + r_2 - 2} b^*_{m, \{r_1, r_2)\}} b_{m', \{r_1, r_2\}} \bra{\phi_0} (A^{m})^{\dagger} \hat{O} A^{m'} \ket{\phi_0}\,,
\end{align}

where
\begin{align}
    b_{m', \{r_1, r_2\}} &= \mathop{\sum_{l_2=0}^{r_2-1} \sum_{l_1=0}^{r_1-1}}_{l_2 + l_1 = m'}  c^{\text{opt}}_{r_2, l_2}  c^{\text{opt}}_{r_1, l_1}\,,
\end{align}
and $b_{m', \{r_1, r_2\}}^*$ denotes the complex conjugate.
By considering computing an observable of the output of another PQSE iteration, and grouping inner products as in Eq.~(\ref{eq:obs_inner_prod_group}), we learn the rule to update $b_{m', \{r_1, r_2\}}$

\begin{align}
     \bra{\psi_3} \hat{O} \ket{\psi_3} &= \sum_{k_3,l_3=0}^{r_3-1} c^{\text{opt}, *}_{r_3, k_3} c^{\text{opt}}_{r_3, l_3} \bra{\psi_2} (A^{k_3})^{\dagger} \hat{O} A^{l_3} \ket{\psi_2}\,, \\
     &= \sum_{k_3,l_3=0}^{r_3-1} c^{\text{opt}, *}_{r_3, k_3} c^{\text{opt}}_{r_3, l_3} \sum_{k_2,l_2=0}^{r_2-1} \sum_{k_1,l_1=0}^{r_1-1} c^{\text{opt}, *}_{r_2, k_2} c^{\text{opt}, *}_{r_1, k_1}c^{\text{opt}}_{r_2, l_2}  c^{\text{opt}}_{r_1, l_1} \bra{\phi_0} (A^{k_1 + k_2 + k_3})^{\dagger} \hat{O} A^{l_1 + l_2 + l_3} \ket{\phi_0}\,.\\
     &= \sum_{k_3,l_3=0}^{r_3-1} c^{\text{opt}, *}_{r_3, k_3} c^{\text{opt}}_{r_3, l_3} \sum_{m, m' = 0}^{r_1 + r_2 - 2} b^*_{m, \{r_1, r_2\}} b_{m', \{r_1, r_2\}} \bra{\phi_0}  (A^{m + k_3})^{\dagger} \hat{O} A^{m' + l_3} \ket{\phi_0}\,.\\
     &= \sum_{m'', m''' = 0}^{r_1 + r_2 + r_3 - 3} \left( \sum_{k_3=0}^{r_3-1} \sum_{m = 0}^{r_1 + r_2 - 2} c^{\text{opt}, *}_{k_3} b^*_{m, \{r_1, r_2\}}\right) \left(\sum_{l_3=0}^{r_3-1} \sum_{m' = 0}^{r_1 + r_2 - 2} c^{\text{opt}}_{l_3}  b_{m', \{r_1, r_2\}} \right)  \bra{\phi_0}  (A^{m''})^{\dagger} \hat{O} A^{m'''} \ket{\phi_0}\,.\\
     &= \sum_{m'', m''' = 0}^{r_1 + r_2 + r_3 - 3} b^*_{m'', \{r_1, r_2, r_3\}} b_{m''', \{r_1, r_2, r_3\}}  \bra{\phi_0} (A^{m''})^{\dagger} \hat{O} A^{m'''} \ket{\phi_0}\,,
\end{align}

where
\begin{align} \label{eq:recon_upd_rule_3rd_it}
    b_{m'', \{r_1, r_2, r_3\}} = \sum_{k_3=0}^{r_3-1} \sum_{m' = 0}^{r_1 + r_2 - 2} c^{\text{opt}}_{r_3, k_3} b_{m, \{r_1, r_2\}}\,.
\end{align}
In this way we can reconstruct properties of the PQSE output state in terms of properties of the initial Krylov basis states.
The rule in Eq.~(\ref{eq:recon_upd_rule_3rd_it}) generalises to the $P$th PQSE iteration as
\begin{align} \label{eq:recon_upd_rule}
    b_{m'', \mathcal{R}'} = \sum_{k=0}^{r_P-1} \sum_{m = 0}^{O(\mathcal{R}) - 1} c^{\text{opt}}_{r_P, k} b_{m, \mathcal{R}}\,,
\end{align}
where $\mathcal{R}' = \{r_1, r_2, \dots, r_P\}$ and $\mathcal{R} = \{r_1, r_2, \dots, r_{P-1}\}$ and $O(\mathcal{R}) = 1 - (P-1) + \sum_{k=1}^{P-1} r_k$ as introduced in the main text in Sec.~(\ref{PQSE}). By iteratively constructing the coefficients $b_{m, \mathcal{R}}$ using the new optimal QSE coefficients $\textbf{c}^{\text{opt}}_{r_P}[\ket{\psi_{P-1}}]$ at each iteration we are able to compute the observable of the PQSE output state at the $P$th iteration as

\begin{align} \label{eq:gen_pqse_ele}
     \bra{\psi_P} \hat{O} \ket{\psi_P} &= \sum_{m, m' = 0}^{O(\mathcal{R}) - 1} b^*_{m, \mathcal{R}} b_{m', \mathcal{R}}  \bra{\phi_0} (A^{m})^{\dagger} \hat{O} A^{m'} \ket{\phi_0}\,.
\end{align}

This also shows how to construct new subspace matrix elements at a new PQSE iteration.
Consider the subspace Hamiltonian at the $P$th iteration. This matrix has elements $H^{(P)}_{ij} = \braket{\psi_{P}\vert (A^{i})^{\dagger} H  A^j \vert \psi_{P}}$ which we can evaluate using Eq.~(\ref{eq:gen_pqse_ele}) with $ \hat{O} = (A^{i})^{\dagger} H  A^j$.
For $S^{(P)}_{ij}$ we would take $\hat{O} = (A^{i})^{\dagger} A^j$.

\newpage

\section{Subspace Dimensions}
\label{app:dim}
In order to aid with interpretation of our numerical results, we provide supporting data of how Krylov basis dimensions change following attempts to improve numerical stability.
We show how many dimensions are discarded after thresholding is applied to the noisy subspace matrices and provide an overview of the typical sequence structure returned by PQSE.
This data is for the spin ring example presented in the main text with a Hamiltonian power generated Krylov basis and a noise strength of $\delta = 10^{-6}$.

\subsection{Threshold quantum subspace expansion}
\begin{figure*}
    \centering
    \begin{tabular}{c|c}
       \includegraphics[width=0.5\textwidth]{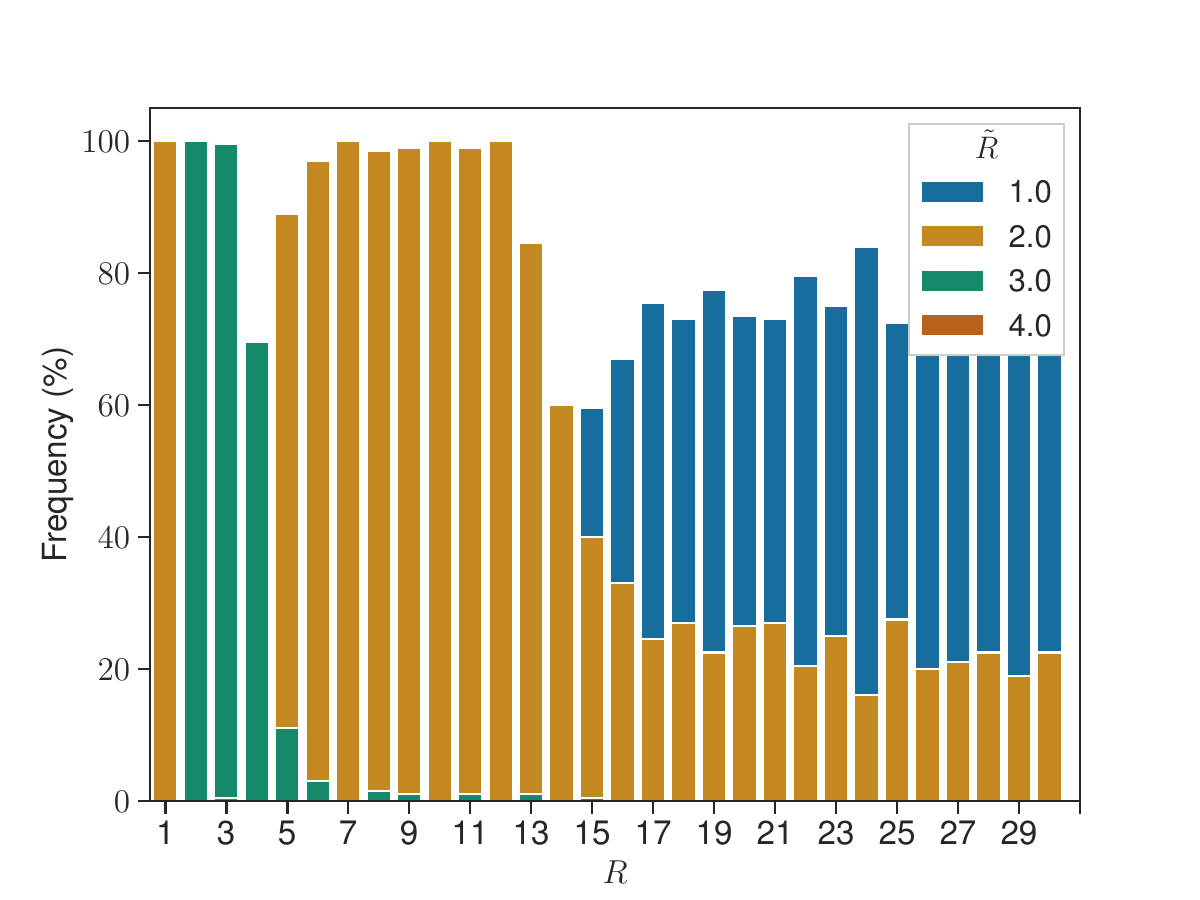}  &  \includegraphics[width=0.5\textwidth]{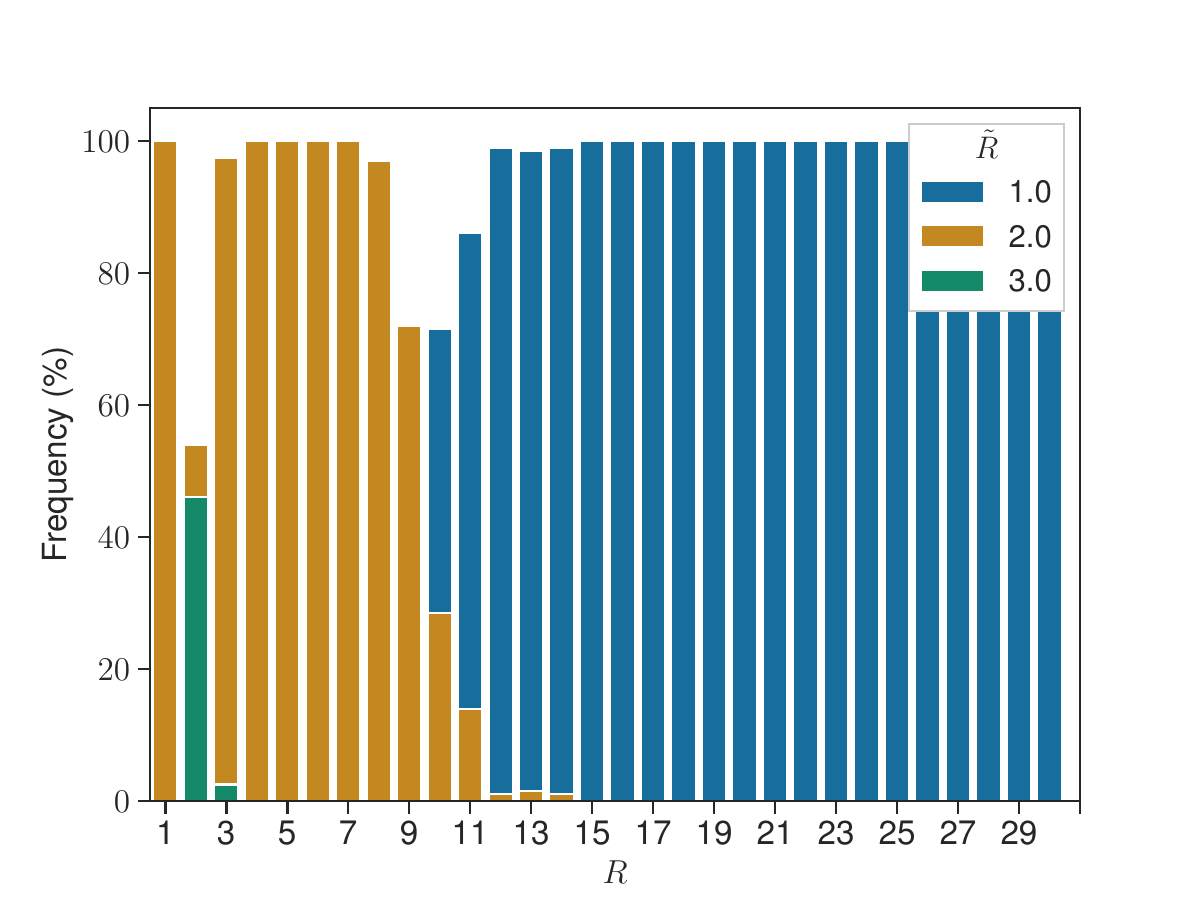} \\
         \includegraphics[width=0.5\textwidth]{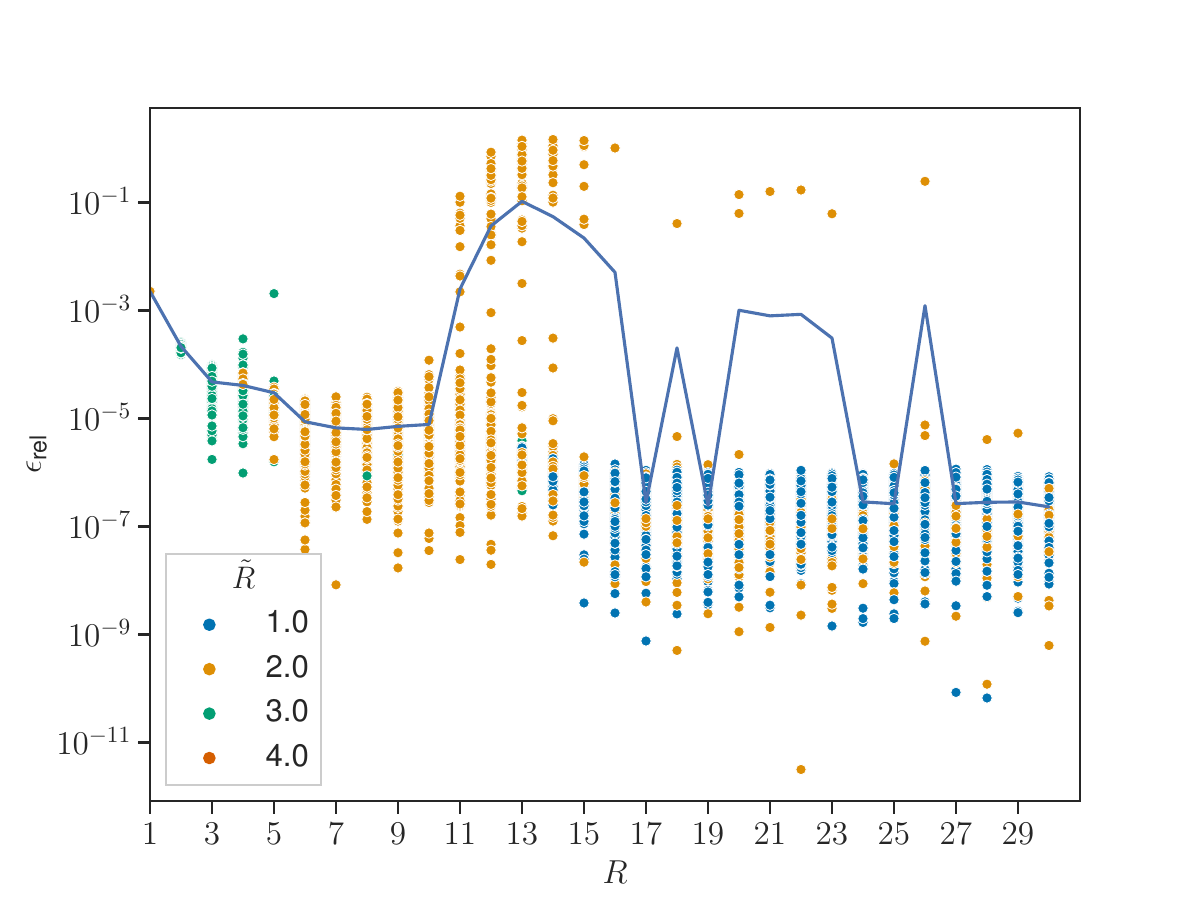}  &  \includegraphics[width=0.5\textwidth]{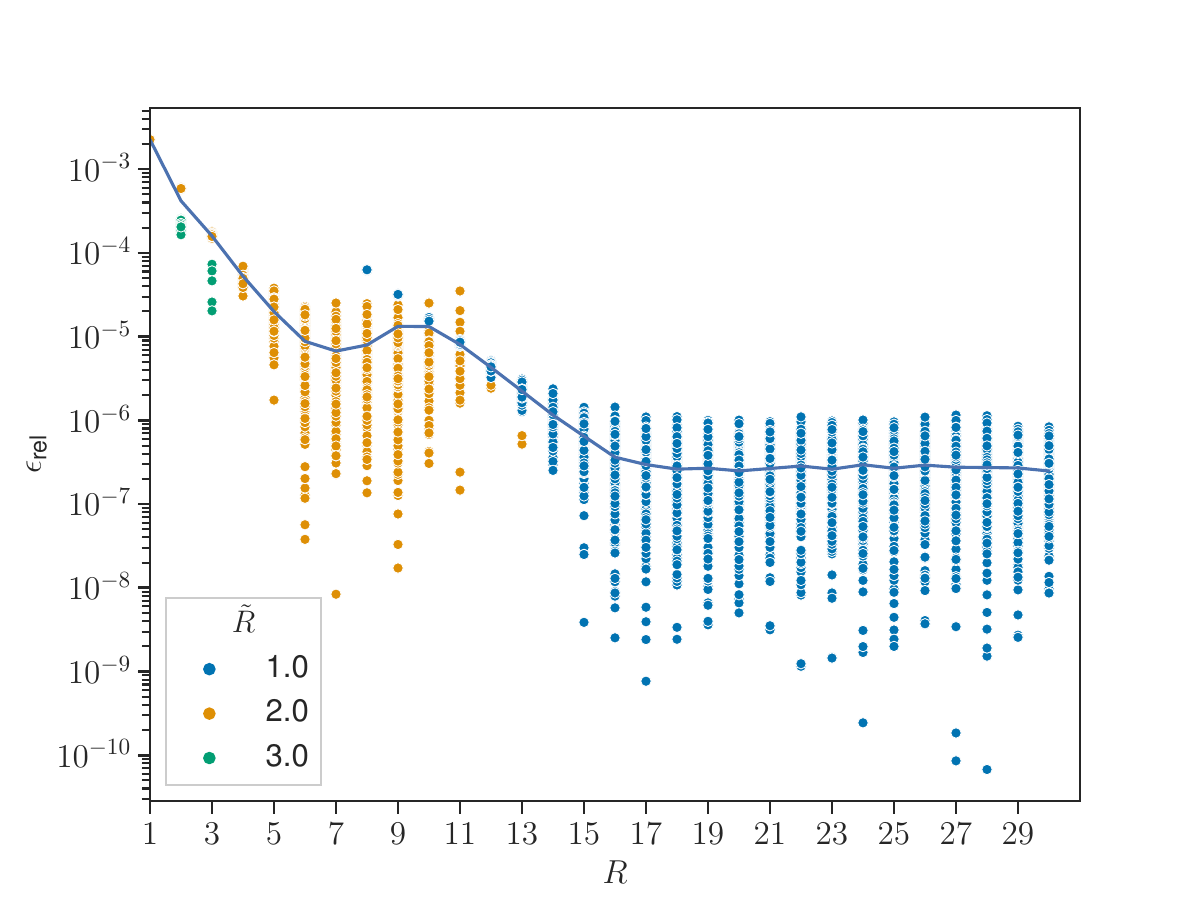}
    \end{tabular}
    \caption{For thresholded QSE (TQSE), the frequency of different post-thresholding basis sizes $\tilde{R}$ (top) and the relative energy error $\epsilon_{\text{rel}}$ (bottom) for increasing initial Krylov basis dimension $R$ for threshold scaling parameters 1.81 (left) and 0.379 (right).    
    For a 10 qubit periodic disordered Heisenberg model with $J$ = 0.1, $h$ = 1, Gaussian noise strength $\delta = 10^{-6}$ and a Krylov basis generated by Hamiltonian powers. Solid line in the bottom plots is the average relative energy error.}
    \label{fig:tqse_dim}
\end{figure*}

The results for TQSE are presented in Fig.~\ref{fig:tqse_dim}, where we show how frequently different post-thresholding basis sizes $\tilde{R}$ appear for a target Krylov basis order $R$, as well as the error associated with these different values of $\tilde{R}$.
We include data for two TQSE threshold values, defined by scaling parameters $a_1 = 1.81$ and $a_2 = 0.379$. Both give a comparable minimum average error but, on average, the experiments with $a_1$ converge much less smoothly than those with $a_2$. We are interested in seeing if this difference in convergence could be explained in terms of the number of dimensions being discarded. 
We see that the majority of subspace dimensions are discarded, with $\tilde{R} \leq 4$ with $a_1$ and $\tilde{R} \leq 3$ with $a_2$. Additionally, with $a_2$ we see that by $R=12$, nearly all experiment instances involve all but $\tilde{R} = 1$ dimension being discarded. For $a_1$, $\tilde{R} = 2$ dimensions are retained in approximately a quarter of the experiments. 
We see that these $\tilde{R} = 2$ experiments for $a_1$ have significantly larger output error than other instances. This suggests the large upward jumps in TQSE error were due to a subset of experiment runs where the TQSE threshold was too large to eliminate spurious eigenvalues.

\subsection{Partitioned quantum subspace expansion}
\begin{figure*}
    \centering
    \includegraphics[width=\linewidth]{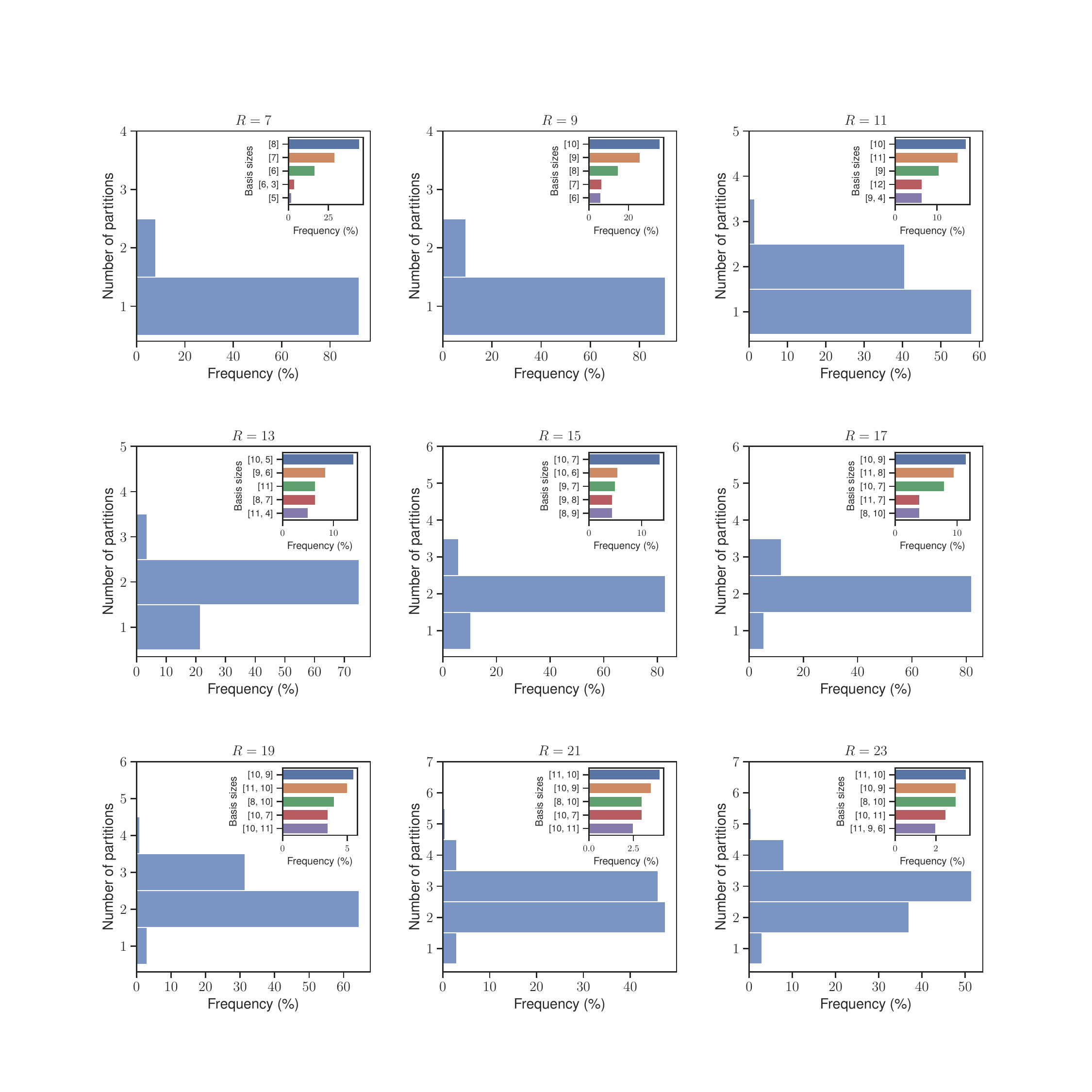}
    \caption{For partitioned QSE (PQSE), the frequency of different partition numbers returned by the algorithm for increasing initial Krylov basis dimension $R$.
    The inset for each $R$ shows top 5 most frequent basis size choices in each partitioning. 
    For a 10 qubit periodic disordered Heisenberg model with $J$ = 0.1, $h$ = 1, Gaussian noise strength $\delta = 10^{-6}$ and a Krylov basis generated by Hamiltonian powers.}
    \label{fig:pqse_basis}
\end{figure*}

We present the frequency of the number of partitions and top five most frequent basis choices selected by PQSE at increasing target Krylov basis order $7 \leq R \leq 23$ in Fig.~\ref{fig:pqse_basis}.
The corresponding energy error for this noise strength can be seen in Fig.~\ref{fig:err_decay}(c) in the main text. 
We find that at small $R$ PQSE will typically choose a single step at the target Krylov basis order, either recreating the output of QSE or discarding a small number of dimensions. 
As $R$ increases PQSE begins to return 2 step sequences more often. At $R \geq 21$, where the algorithm appears to converge in energy error, 2 or 3 step sequences are most frequently selected.
Additionally, as $R$ increases and the number of possible sequence combination increases, PQSE returns a broader range of possible sequences. Some partitionings appear to be selected repeatedly, suggesting convergence to a fixed set of equally well performing sequences.

\newpage

\end{document}